\documentclass[aps,amsfonts,nofootinbib,twocolumn,a4paper]{revtex4}
\usepackage{graphicx}
\usepackage{amsmath}
\def\be{\begin{equation}}
\def\ee{\end{equation}}
\def\bea{\begin{eqnarray}}
\def\eea{\end{eqnarray}}

\begin{document}

\title{Cosmological constraints combining $H(z)$, CMB shift  and SNIa observational data}

\author{Ruth Lazkoz$^1$ and Elisabetta Majerotto$^2$}

\affiliation{\vspace*{0.2cm} $^1$Fisika Teorikoa, Euskal Herriko
Unibertsitatea, 48080~Bilbao, Spain\\
$^2$Institute of Cosmology \& Gravitation, University of
Portsmouth, Portsmouth~PO1~2EG, UK\\
E-mail: ruth.lazkoz@ehu.es and elisabetta.majerotto@port.ac.uk}

\date{\today}

\begin{abstract}

Recently $H(z)$ data obtained from differential ages of galaxies have been proposed as a new geometrical probe of dark energy. In this paper we use those data, combined with other background tests (CMB shift  and SNIa  data), to constrain  a set of %popular  
 general relativistic %parametrizations of $H(z)$
dark energy models
 together with some  other models motivated by extra dimensions. 
 Our analysis rests mostly on Bayesian statistics, and we conclude that  LCDM is at least substantially  favoured, and that extradimensional models are less favoured
than general relativistic ones.

\end{abstract}

\maketitle

\section{Introduction}

%General introduction about accelerating Universe

%Our work:
%         choice of models
%         choice of data
%         choice of method
%         possible problems
The almost undisrupted flow of data potentially useful for cosmological model selection and parameter fitting provides day by day a clearer picture of the evolution of the Universe, which according to a wide consensus seems to be currently accelerating. Yet it is uncertain how strong the conclusions of these sorts of investigation are, as there are not only statistical uncertainties 
to worry about, but possible theoretical biasing in the tests used. 

This being the situation, the community is eager to be able to devise new tests 
and/or improve the understanding of the shortcomings of the existing ones towards
refining them. The underlying hope is that, at least from a phenomenological perspective, cosmologists will eventually be able to
tell with high precision how fast the Universe is expanding at present, how long this speed up has lasted, and how the acceleration rate has changed over the recent past.

 Recently Simon et al. \cite{simon} have published Hubble function $H(z)$ data extracted from differential ages of passively evolving galaxies. The use of these data to constrain the background evolution of the Universe is interesting for several reasons. First, it can be used together with other cosmological tests in order to get useful consistency checks or tighter constraints on models. Second, in contrast to standard candle luminosity distances or standard ruler angular diameter distances, the Hubble function is not integrated over. This will allow, mostly in the future, when the systematic errors associated with the cluster physics have been reduced, to get cleaner constraints on the cosmological parameters. 
The same data have also been recently used in \cite{ratra, use hubble} to constrain several cosmological models.
In this paper we use them in combination with the latest SNe Ia set compiled by Davis et al. \cite{Davis} and the CMB shift parameter as calculated by Wang and Mukherjee \cite{Wang} from WMAP 3 \cite{Spergel}.
 In contrast to other works, we have chosen not to use the baryon acoustic oscillations (BAO) data because, as pointed out in \cite{roy}, there is a level of uncertainty in the use of the measure $A$ given in \cite{eisenstein} to test non-LCDM models.% During the completion of this work (which has taken several months due to
%heavy calculations), we have become aware of some couple of papers which question the validity of  the use of the CMB shift parameter alone \cite{multamaki,wright,wang07}, but as they state the effect are probably tiny in parametrizations where dark energy has a negligible contribution at high redshifts.

We have chosen to put constraints on 6 different cosmological models that show late-time acceleration (see e.g. \cite{Riess1998, Riess:2004nr, riess07}). For all of them we have assumed  flat space, consistent with the inflationary prediction that the curvature density parameter  $\Omega_k \sim 0$ \cite{inflation} and with the results obtained by CMB experiments \cite{flat cmb, Spergel}. 
%(though relaxing the LCDM underlying hypothesis one may obtain different results for the curvature content of the Universe, as pointed out in \cite{marina})

 Three of the models are dark energy models while in the remaining three models the acceleration arises from or is modified by a five-dimensional modification of gravity motivated by extra dimensional physics. One also can group these models by pairs according to the number of parameters apart from $H_0$. LCDM and DGP \cite{dgp, deffayet} have only one parameter, i.e. the matter density parameter. QCDM, i.e. dark energy parametrized as a perfect fluid with a constant equation of state, and LDGP, a non self-accelerating solution for the DGP action \cite{ldgp}, add to the matter density one further parameter: the equation of state and the crossover scale (i.e. the scale at which the 5-dimensional behaviour becomes effective), respectively. The last two models chosen are the Chevalier-Polarski-Linder parametrization \cite{cp, linder} (which contains two parameters characterizing the evolution of the equation of state) and the QDGP model \cite{qdgp paper}, a phenomenological extension of the LDGP model, where the cosmological constant is replaced by a dark energy fluid with a constant equation of state. Similar analyses of a subset of these models  using recent datasets partially overlapping with ours have been done in \cite{similar_papers, dgp_goobar, other_constraints} and our work may be complementary to that.

The plan of the paper is as follows. In Sec. \ref{models} we outline the models to be investigated and the corresponding forms of $H(z)$. Then, in Sec. \ref{obs} we give details of the three datasets considered, and we explain how they can be used to constrain the background of cosmological models. After that,  in Sec. \ref{estsec} we review the concepts and techniques we resort to in our statistical analysis of the data. Finally, we present our results and conclusions in Sec. \ref{rescon}  (accompanied by some additional statistical information and considerations about the effective equations of state in the different models in the Appendices).

\section{Models and Parametrizations}\label{models}
%Some of the models we are going to consider will have unphysical regions, i.e. regions of the parameter space where one could have $H^2$ for some  %values of redshifts. This can lead to divergences in the luminosity function, so in we will try to cure those problems. For some parametrizations we will discard completely such region of the space, so no divergences exist at any redshift, whereas for some other parametrizations others we will just 
%demand the divergence occurs to the future, i.e. for redshifts larger than $z=0$. This will be done in the exceptional cases where 
%demanding  that no divergence appears at no redshift occurs only a measure-zero region of the parameter space. Specifically, we will assign in our %codes a negligible probability to those pathological cases.

As discussed in detail in \cite{sahni}, extracting information about the properties of dark energy from a reconstruction of $H(z)$ is not in principle problematic from the mathematical perspective, but it is hard in practice because one only has at hand a discrete set of noisy data (for several redshifts) from which one
wishes to infer information valid for all redshift values within a given interval. %The most common way to 
%get across this obstacle is  % to make an Ansatz on either $D_l(z)$ \cite{dl},
%$w_{de}(z)$ \cite{wans,perivol,cp,linder}, or directly on $H(z)$ \cite{Nesseris:2005ur, geom, deffayet, ldgp, qdgp paper}.
%eiher to make an Ansatz on the luminosity distance $D_L(z)$ \cite{dl} or to derive $w_{de}(z)$ \cite{wans,perivol,cp,linder} and $H(z)$ \cite{Nesseris:2005ur, geom, deffayet, ldgp, qdgp paper} from a cosmological model.
The various approaches to confronting models with data are based on deriving the luminosity distance $D_L(z)$ \cite{dl}, $w_{de}(z)$ \cite{wans,perivol,cp,linder} or $H(z)$ \cite{Nesseris:2005ur, geom, deffayet, ldgp, qdgp paper}. 

%Here we will present observational constraints on Ans{\"a}tze of the two last class of those we have just mentioned. In particular, as already mentioned in the Introduction, we will compare six models altogether:
Here we will use the second approach and compare to observations six cosmological models altogether:
 the models can be grouped in pairs according to the number of parameters (one, two or three); in addition each pair will contain one %model arising from 
%a guess on $w_{de}(z)$, in the dark energy interpretation spirit, and another model responding to an Ansatz on $H(z)$, in the dark gravity fashion.
general relativistic dark energy model and one  model
%In particular these will be scenarios
inspired by extradimensional modifications to gravity.%, i.e. cosmologies in which dark energy will be a geometrical illusion, but with manifest effects.
%where dark energy will be a geometrical illusion, but with manifest effects.

\subsection{LCDM %Ansatz
}
In general, %if one sets off from
if we consider a general relativistic  model of dark energy  with an equation of state that depends on the redshift, $w_{de}(z)$, the expression for $H^2(z)$
obeys
%\begin{widetext}
\begin{eqnarray}\nonumber
\frac{H^2(z)}{H_0^2}&=&\Omega_m(1+z)^3\\
&&+(1-\Omega_m)\exp\left[3\int_0^z\frac{1+w_{de}(x)}{1+x}dx\right],
\end{eqnarray}
%\end{widetext}
where one assumes that in addition to dark energy the Universe contains dust (dark matter and baryons).

The LCDM %Ansatz 
model
corresponds to the choice $w_{de}(z)=-1$, i.e. dark energy is a cosmological constant $\Lambda$. Thus
\begin{equation}
\frac{H^2(z)}{H_0^2}=\Omega_m(1+z)^3+1-\Omega_m,
\end{equation}
and the free parameter of the model is $\Omega_m$.
%Considering current observational odds this Ansatz remains admissible despite it poses an important theoretical problem. 
LCDM is consistent with all data, but there is a theoretical problem in explaining the observed value of $\Lambda$.
%%%%%%%%%%%%%%%%%%%%%%%%%%%%%%%%%%%%%%%%%%%%%%%%%%%%%%%%%%%%%%%%%%%%%%%%%%%
\subsection{DGP% Ansatz
}
The %Ansatz 
dark gravity model
inspired by the Dvali-Gabadadze-Porrati (DGP) braneoworld model \cite{dgp}
 was given by Deffayet \cite{deffayet},
and it represents a simple alternative to the standard LCDM cosmology,
with the same number of parameters.

In this model the late Universe self-accelerates, not because of dark energy, which is absent, but rather 
due to an infrared modification of gravity. Explicitly one has
\begin{equation}
\frac{H(z)}{H_0}=\frac{1-\Omega_m}{2}+\sqrt{\frac{(1-\Omega_m)^2}{4}+\Omega_m(1+z)^3}.
\end{equation}

%In this model , as in any other Ansatz where one starts off by postulating theoretically or deriving the functional dependence of $H(z)$, the effective dark energy equation of state parameter $w_{\rm eff}$ could be read off from
One can define an effective dark energy equation of state $w_{\rm eff}$ by imposing 
$H^2=H_0^2\Omega_m(1+z)^3+8\pi G \rho_{\rm eff}/3$ along with
$\dot{\rho}_{\rm eff}+3H(1+w_{\rm eff})\rho_{\rm eff}=0 $ to obtain, implicitly
\begin{equation}
 w_{\rm eff}=\frac{\displaystyle\frac{2}{3}(1+z)\frac{d \ln E(z)}{dz}-1}{1-E^{-2}(z)\Omega_m(1+z)^3}.\label{weff},
\end{equation}
where $E(z)\equiv H(z)/H_0$, and explicitly
\begin{equation}
w_{\rm eff}(z)=\frac{\Omega_m-1-\sqrt{(1-\Omega_m)^2+4\Omega_m(1+z)^3}}{
2\sqrt{(1-\Omega_m)^2+4\Omega_m(1+z)^3}}\, .
 \end{equation}
For this scenario one can see that $\lim_{z\to-1}w_{\rm eff}=-1$, so the final asymptotic state is a de Sitter model.
\subsection{QCDM %Ansatz
}
This model arises from the simplest generalization of %the LCDM Ansatz,
LCDM,
 which consists in taking a constant value of $
w_{de}=w$ different, in general, from $-1$, so one gets
\begin{equation}
\frac{H^2(z)}{H_0^2}=\Omega_m(1+z)^3+(1-\Omega_m)(1+z)^{3(1+w)}.
\end{equation}
Despite its simplicity, this %Ansatz 
model can prove useful, as a preference for this parametrization over the LCDM case
will provide support for the evolutionary nature of dark energy. %This seems to be, indeed  the case, according to the result in \cite{riess07},  where it was shown that in the redshift range of interest there ($0.2<z<1.2$) the fit was statistically better than that of LCDM.

\subsection{LDGP %Ansatz
}
This %Ansatz 
model
%\cite{ldgp} stems essentially from the same modification of the Friedman equation as the DGP Ansatz reported above, but there are two crucial differences.  The first difference is that although the two models are derived from a modification to gravity that introduces a linear term in $H$ in the Friedmann equation, for the LDGP Ansatz the (customary) quadratic and the (novel) linear term in $H$ have the same sign, whereas
%for the so called DGP Ansatz the signs are opposite. The second crucial difference is that one considers that the observed universe (which is a braneworld) has a homogeneous distribution of not only
%pressureless matter, but also a cosmological constant which is responsible for the late-time acceleration. In contrast, in the DGP Ansatz the cosmological constant is absent. Nevertheless, considering the existence of vacuum energy pays off in some respects, because for instance it allows for phantom behaviour without phantom dark energy.
represents the non self-accelerating branch of DGP (the two separate branches of DGP arise from the two possible ways to embed the 4D brane universe in the 5D spacetime; see also \cite{ldgp, lazmaamaj} for further explanation). To generate acceleration a cosmological constant is needed. There are two main features of this model. One is the screening effect on $\Lambda$ due to the presence of the extra dimension, that allows for a higher  value of the cosmological constant. The other is the possibility of having effective phantom behaviour without any phantom field and any of the associated instabilities. In addition, this model does not have ghosts, unlike the self-accelerating DGP.

The evolution of the model  is governed by the relation
\begin{equation}
\frac{H(z)}{H_0}=\sqrt{\Omega_m(1+z)^3+\Omega_{\Lambda}
}-\sqrt{\Omega_{r_c}},
\end{equation}
where  on the one hand $\Omega_{\Lambda}=1+2\sqrt{\Omega_{r_c}}-\Omega_m$ for the requirement of flatness (see \cite{lazmaamaj}), and on the other hand the parameter $\Omega_{r_c}$ is related to the crossover scale which signals the transition from the general relativistic to the modified gravity regime (see again \cite{ldgp} for details). 
Observational constraints on this particular model using SN, CMB shift and BAO data were studied in \cite{lazmaamaj}, where it was found that statistically the best fit corresponded to the LCDM limit of the model.

The same procedure to obtain $w_{\rm eff}$ that we mentioned in the DGP model applies here, so one will have, from Eq. (\ref{weff}),
\begin{equation}
w_{\rm eff}(z)=-1-\frac{\sqrt{\Omega_{r_c}}\Omega_m (1+z)^3}{[\Omega_{\Lambda}-2\sqrt{\Omega_{r_c}}E(z)][\sqrt{\Omega_{r_c}}+E(z)]}.
\end{equation}

\subsection{Chevallier-Polarski-Linder Ansatz}
This is a %very popular 
widely used
Ansatz, which generalizes %the QCDM Ansatz,
QCDM
 to %display better  adaptability to evolutionary dark energy. 
allow for evolution in the dark energy equation of state, as required by most realistic scalar field models. 
(It was first discussed in \cite{cp}, and reintroduced independently in ·\cite{linder}.)

 Explicitly it corresponds to $w_{\rm de}(z)=w_0 + w_1 (1-(1+z)^{-1})$ and therefore

%\begin{widetext}
\begin{eqnarray}\nonumber
\frac{H^2(z)}{H_0^2}&=&\Omega_m(1+z)^3\\
&&+(1-\Omega_m)(1+z)^{3(1+w_0+w_1)}e^{-3 \frac{w_1 z}{1+z}}.
\end{eqnarray}
%\end{widetext}
Among its desirable features two stand out: first, this parametrization of dark energy remains finite at large redshifts, and second, the physical interpretation of the model is simple as the parameter $w_1$ is a measure of the scalar field potential slow roll factor $V'/V$  in the case of quintessence \cite{linder}.

\subsection{QDGP %Ansatz
}
This %parametrisation is a phenomenological variation of the LDGP Ansatz which was introduced in \cite{qdgp paper}
%\begin{widetext}
model, introduced in \cite{qdgp paper}, is a generalization of LDGP where the cosmological constant is replaced by dark energy with a constant equation of state $w$, in general different from $-1$, that represents a new parameter of the model.
The modified 4D Friedman equation has the following form:
\begin{equation}\
\frac{H(z)}{H_0}=\sqrt{\Omega_m(1+z)^3+\Omega_w(z)+\Omega_{r_c}}-\sqrt{\Omega_{r_c}}
\end{equation}
where
$\Omega_w(z)\equiv ( 1+2\sqrt{\Omega_{r_c}}-\Omega_m)(1+z)^{3(1+w)}$.
%\end{widetext}
%In this case one postulates that the dark matter is accompanied by quiessence, rather than a cosmological constant, so a new parameter enters the game. %, and perhaps it could lead to a better fit to the data than its precursor, the LDGP model.
Clearly, in this case too one can make use of Eq. (\ref{weff}) to derive $w_{\rm eff}$:
\begin{equation}
w_{\rm eff} = -1 +\frac{(1+w)\Omega_w(z)E(z)-\sqrt{\Omega_{r_c}}\Omega_m(1+z)^3}{[E(z)+\sqrt{\Omega_{r_c}}][\Omega_w (z)-2E(z)\sqrt{\Omega_{r_c}}]}.
\end{equation}  

\section{Observational tests}\label{obs}
\subsection{Hubble parameter observations}
The Hubble parameter depends on the differential age of the Universe in terms of redshift, specifically
\begin{equation}
H(z)=-\frac{1}{1+z}\frac{dz}{dt}.
\end{equation}
Thus, measuring $dt/dz$ allows us to determine $H(z)$. As described first in \cite{jimenez_loeb} and in \cite{stern, simon}, it is possible to use absolute ages of passively evolving galaxies to  compute  values of $dt/dz$. The galaxy spectral data used by \cite{simon} 
come from the Gemini Deep Deep Survey \cite{gdds} and archival data \cite{archival}.
In broad terms, the authors of these references bin together galaxies with a redshift separation which is small enough so that the galaxies in the bin have roughly the same age; then, they calculate age differences between bins which have a small age difference which is at the same time  larger than the error in the age itself \cite{simon}. The outcome of this process is a set of 9 values of the Hubble parameter versus  redshift (see Table \ref{tab H}). A particularly nice feature of this test is that differential ages are less sensitive to systematic errors
than absolute ages \cite{dunlop}.

\begin{table}[htp!]\caption{H(z) data from \cite{simon} (in units of ${\rm \, km\, s}^{-1}{\rm Mpc}^{-1}$)}
%\tablehead{\colhead{$z$} & \colhead{$H(z)^{\rm a}$} \\ \colhead{} & \colhead{($\kmsmpc$)}}
%\startdata
\begin{center}
\begin{tabular}{l@{\hspace{0.2in}}l@{\hspace{0.2in}}l}
\hline
$z$ & $H(z)$ & $\sigma$\\
\hline
0.09 & 69 & 12\\
0.17 & 83 & 8.3\\
0.27 & 70 & 14\\
0.4 & 87 & 17.4\\
0.88 & 117 & 23.4\\
1.3 & 168 & 13.4\\
1.43 & 177 & 14.2\\
1.53 & 140 & 14\\
1.75 & 202 & 40.4\\
\hline
\end{tabular}\label{tab H}
\end{center}
\end{table}

Observed values of $H(z)$ can be used to place constraints on different models  of the expansion history of the Universe by minimizing
the quantity
 \begin{eqnarray}
\chi^2_{\rm H}(H_0,\{\theta_i\})=\sum_{j=1}^{9}\frac{(H(z_j;\{\theta_i\})-H_{\rm obs}(z_j))^2}{\sigma_{\rm H,j}^2}.
\end{eqnarray}

This test has already been used to constrain several cosmological models in \cite{ratra, use hubble}, and it does not seem to provide tight constraints on its own, so it seems necessary to combine it with other tests.
%one of the goals of this paper is to investigate whether combination with other tests brings some improvement.
Now, since one of the other tests we are using (the CMB shift) does not constrain $H_0$ we will regard this parameter as a nuisance one, and so it will be more convenient to marginalize over it (see for instance \cite{rob}) and work with the quantity
\begin{eqnarray}
 \hat\chi^2_{\rm H}({\{\theta_i\}})=-2\log\left(\int \pi_H(H_0) e^{-\chi^2_{H}(H_0,\{\theta_i\})}
dH_0\right).\end{eqnarray}
The quantity $\pi_H(H_0)$ is a so called prior probability function (see section \ref{estsec}) which reflects some previous knowledge about preferred values of $H_0$.  We have independent evidence from the Chandra X-ray observatory giving $H_0=77\pm4\; {\rm km} {\rm s}^{-1} {\rm Mpc}^{-1}$
\cite{bonamente}. This result is  consistent with the 3 year WMAP dataset result  $H_0=73\pm3\; {\rm km} {\rm s}^{-1} {\rm Mpc}^{-1}$
\cite{Spergel}
and with the Hubble Space Telescope key project result  $H_0=73\pm8\; {\rm km} {\rm s}^{-1} {\rm Mpc}^{-1}$
\cite{freedman} . Here we choose to use the Chandra X-ray result (for statistical independence with respect to the other tests to be used later) and use this result to place a Gaussian prior on $H_0$ so that
\begin{equation}
\pi_H(H_0)=\frac{1}{\sqrt{2\pi}\sigma_{H_0}}e^{-(H_0-H_0^{obs})^2/2\sigma^2_{H_0}}
\end{equation}
where obviously $H_0^{obs}=77$ and $\sigma_{H_0}=4$.

\subsection{The Davis et al. 2007 dataset}
This dataset is one of the latest supernovae catalogs to be completed. It consists of 192
SNe classified as type Ia up to a redshift of $z=1.755$
\cite{Davis}. The authors combine the set compiled by \cite{Wood-Vasey}  and the 30 new SNe data at high redshift ($0.216 \leq z \leq 1.755$) recently discovered with
the Hubble Space Telescope (HST) \cite{riess07}. The first of those two sets consists of the addition of three smaller datasets. The first subset is made of the SNe Ia data from the ESSENCE project \cite{Wood-Vasey}, a ground-based survey designed to detect about 200 SNe Ia in the range $z=0.2-0.8$. The second is made of the Supernova Legacy Survey (SNLS) \cite{Astier} data. These have been refitted by \cite{Wood-Vasey} with the same lightcurve fitter used for the ESSENCE data. The third subset corresponds to the nearby SNe already presented in \cite{Hamuy, Riess1998, Jha} as refitted and used by \cite{Wood-Vasey}. As regards the HST SNe, it has been necessary to perform a normalization since the refitting of the HST lightcurves with the fitter used for the ESSENCE data is still in progress. The normalization has been done using the low redshift sample common to both sets and the error in the normalization is included in the distance errors for the HST SNe. Summarizing, the Davis et al. dataset consists of 60 ESSENCE supernovae, 57 SNLS supernovae, 45 nearby supernovae and 30 HST supernovae. It is
available at

\noindent \texttt{http://www.dark-cosmology.dk/archive/SN,}
\texttt{http://braeburn.pha.jhu.edu/\~{}ariess/R06}
 and 
\texttt{http://www.ctio.noao.edu/essence.}

The statistical best fits of theoretical forms of $H(z)$ using supernovae data rest on the definition for the distance modulus
 \begin{equation}
 \mu_{\rm th}(z_i)=5\log_{10}(d_L(z;\{\theta_i\}))+\mu_0\end{equation}
 where 
  \begin{equation}d_L(z;\{\theta_i\})=(1+z)\int_0^z\frac{H_0dz}{H(z;H_0,\{\theta_i\})}\end{equation}
   is the dimensionless luminosity distance which is related to its dimensional counterpart through
   $D_L(z;H_0,\{\theta_i\})=c \,d_L(z;\{\theta_i\})/H_0$.
   
 The best fits are obtained by minimizing the quantity (see Sec. \ref{estsec})
 \begin{eqnarray}
\chi^2_{\rm SN}(\mu_0,\{\theta_i\})=\sum_{j=1}^{192}\frac{(\mu_{\rm th}(z_j;\mu_0,\{\theta_i\})-\mu_{\rm obs}(z_j))^2}{\sigma_{\mu,j}^2},
\end{eqnarray}
where
 the $\sigma_{\mu,j}$ are the measurement variances \cite{Davis}. 
The nuisance parameter $\mu_0$ encodes the Hubble parameter and the absolute magnitude $M$ \cite{goobar}
and has to be marginalized over \cite{rob}. So one will
actually be working with the quantity
\begin{equation}
 \hat\chi^2_{\rm SN}({\{\theta_i\}})=-2\log\left(\int e^{-\chi^2_{SN}(\mu_0,\theta_1,\dots,\theta_n))}
d\mu_0\right).
\end{equation}
A frequently used alternative \cite{min}, consists in minimizing the quantity 
\begin{equation}\tilde \chi^2_{\rm SN}({\{\theta_i\})}=c_1-\frac{c_2^2}{c_3} \end{equation}with respect to the other parameters.
Here\begin{eqnarray}
&&c_1=\sum_{j=1}^{192}\frac{\left(\mu_{\rm th}(z_j,\mu_0=0,{\{\theta_i\}})-\mu_{\rm obs}(z_j)\right)^2}{\sigma_{\mu,j}^2}\quad\\
&&c_2=\sum_{j=1}^{192}\frac{\mu_{\rm th}(z_j,\mu_0=0,{\{\theta_i\}})-\mu_{\rm obs}(z_j)}{\sigma_{\mu,j}^2}\quad\\
&&c_3=\sum_{j=1}^{192}\frac{1}{\sigma_{\mu,j}^2}\quad.
\end{eqnarray}
It is trivial to see $\tilde \chi^2_{\rm SN}$ is just a version of $\chi^2_{\rm SN}$ minimized with respect to $\mu_0$. To that end
it suffices to notice that \cite{min}
\begin{eqnarray}
 \chi^2_{\rm SN}{(\mu_0,\{\theta_i\})}=c_1-2c_2\mu_0+c_3\mu_0^2,
\end{eqnarray}
which clearly becomes minimum for $\mu_0=c_2/c_3$, and so we can see $\tilde \chi^2_{\rm SN}( \{\theta_i\})\equiv\chi^2_{\rm SN}{(\mu_0=0,\{\theta_i\})}$.
Furthermore,
one can check that the difference between $\hat\chi^2_{SN}$ and  $\tilde\chi^2_{SN}$  is negligible and therefore we will 
just use the first of them.

%%%%%%%%%%%%%%%%%%%%%%%%%%%%%%%
%\subsection{The CMB shift}
%%%%%%%%%%%%%%%%%%%%%%%%%%%%%%%
\subsection{The CMB shift}
The CMB shift $R$ is arguably the parameter with least model-dependence among those which can be inferred from CMB data, provided that the dark energy density parameter is negligible at recombination, %because 
and
it does not depend on $H_0$. It is directly proportional to the ratio of the locations of the first peak in the temperature angular power spectrum in the model one wants to characterize ($l_1^{TT}$) and in a reference flat SCDM model 
($l_1^{'TT}$), that is,
\begin{equation}R\equiv2\frac{l_1^{TT}}{l_1^{'TT}}\approx H_0\sqrt{\Omega_m}\int_0^{z_{\rm rec}}\frac{dz}{H(z)}\label{cmbshift},\end{equation}
where $z_{\rm rec}$ is the redshift of recombination (a factor of $2$ has deliberately been introduced  so as to reconcile the different
definitions of the shift parameter in the literature). The approximate expression in the last equality
of (\ref{cmbshift}) is the definition given in  \cite{bet} and when we turn to numerical tests we will take it as exact, i.e. we will ignore  the fact that it is only approximately equal to other definitions. In the following lines we will show how to obtain this approximate expression for $R$.
%%%%%%%%%%%%%%%%%%%%%%%%%%%%%%%
%calculate the CMB shift at first order
%%%%%%%%%%%%%%%%%%%%%%%%%%%%%%%
In an arbitrary  model one has
\begin{equation}
l_1^{TT}=\pi\frac{D_A(z_{\rm rec})}{ r_s(z_{\rm rec})}\label{ldef}
\end{equation}
where the last scattering sound horizon scale $r_s(z_{\rm rec})$
is given by

\begin{equation}r_s(z_{\rm rec})=a_{\rm rec}\int_0^{a_{\rm rec}}\frac{c_s(a)da}{a^2H(a)},\end{equation}
with $c_s$ the sound speed in the model, 
and (under the assumption of flatness) the sound horizon angular diameter distance is in turn given by 
\begin{equation}D_A(z_{\rm rec})=\frac{c}{1+z_{\rm rec}}\int_0^{z_{\rm rec}}\frac{dz}{H(z)}.\label{angdist}\end{equation}
Arguably, one can take $c_s(a)$ to be constant, and in addition we will take $a_{\rm rec}$ and $a_{\rm eq}$ to be model independent.
Approximate expressions will be obtained using $1\gg a_{\rm rec}\gg a_{\rm eq}$. For a fiducial matter-radiation model one gets 
\footnote{We present our expression in a fashion more similar to that in Eq. (10) of Ref \cite{Percival}
although it is completely equivalent to Eq. (18) in  Ref.  \cite{pernes})}
 \begin{eqnarray}r'_s(z_{\rm rec})=&&\frac{a_{\rm rec}}{H_0\sqrt{\Omega'_m}}\int_0^{a_{\rm rec}}\frac{c_sda}{(a+a_{\rm eq})^{1/2}}=\nonumber\\&&
\frac{2 c_s}{H_0
   \sqrt{\Omega' _m}} \left(\sqrt{a_{\rm eq}+a_{\rm rec}}-\sqrt{a_{\rm eq}}\right),
\end{eqnarray}
but if one makes the more stringent assumption that the fiducial model is a SCDM one ($\Omega' _m=1$) then from the latter
it follows that
 \begin{eqnarray}
r'_s(z_{\rm rec})\approx\frac{c_s }{H_0}\,{a^{3/2}_{\rm rec}},\label{shprime}\end{eqnarray}
and
\begin{equation}D'_A\approx\frac{2c }{H_0}\,a_{\rm rec}.\label{angdistprime}\end{equation}
On the other hand, and in an arbitrary  model 
\begin{equation}r_s(z_{\rm rec})\approx
\frac{c_s }{H_0\sqrt{\Omega_m}}\,{a^{3/2}_{\rm rec}},\label{shnoprime}\end{equation}
 so using the definition  in Eq. (\ref{ldef}) combined with Eqs. (\ref{angdist},\ref{shprime},\ref{angdistprime},\ref{shnoprime}),
one obtains Eq. (\ref{cmbshift}).
Constraints on the parametrizations of $H(z)$ using reported values of the CMB shift (based on observations) will be
obtained from  minimization of the quantity (see Sec. \ref{estsec})
 \begin{eqnarray}
\chi^2_{\rm CMB}(\{\theta_i\})=\frac{(R(z_{rec};\{\theta_i\})-R_{\rm obs}(z_{\rm rec}))^2}{\sigma_{\rm R}^2}.
\end{eqnarray}
We use the value $R=1.70\pm0.03$ of the CMB shift parameter calculated  by \cite{Wang} from the WMAP 3 data \cite{Spergel}
(for the computed value $z_{\rm rec}=1090$).
Recently some points regarding the use of the CMB shift parameter have been raised and discussed by \cite{multamaki, wright, wang07} and this is an issue that deserves additional clarification. Furthermore, one must be cautious,  when weighting the conclusions reached
using the CMB shift as it extends the redshift integration out to very
large redshift, so that arbitrary parametrizations called to be applicable for
such large redshift range can be very sensitive to errors.

%%%%%%%%%%%%%%%%%%%%%%%%%%%%%%%
%\section{\label{estsec} Statistics and data analysis}
%%%%%%%%%%%%%%%%%%%%%%%%%%%%%%%
\section{\label{estsec} Statistics and data analysis}

Parameter estimation in the context of a given (cosmological) model depending on some parameters has two ultimate goals: one
is the determination of the ``most likely'' values of the parameters to  yield a series of available observational data,  the other is 
measuring our degree of confidence in the fact that those data where generated by values of those parameters lying in an estimated
interval. A related task is to compare different models using information retrieved in the parameter estimation process. The estimators used
in both these jobs (parameter estimation and model comparison) are different in the two main approaches to statistics (frequentist and Bayesian). In the remainder of this section we give a short account of these topics. 
%%%%%%%%%%%%%%%%%%%%%%%%%%%%%%%
%\subsection{Parameter estimation}
%%%%%%%%%%%%%%%%%%%%%%%%%%%%%%%
\subsection{Parameter estimation}

By definition ${\cal L}(\{d_j\}\vert\{\theta_i\},{\cal M})$ represents the the unnormalized probability density function (aka likelihood) that one measures the data $\{d_j\}$ given the model ${\cal M}$ is true and its parameters take values $\{\theta_i\}$ \cite{jussi}. 
Even though we will try to keep the discussion in this section as general as possible, when we turn to analyze particular datasets we will assume
as customary that the measurements are normally distributed around their true value so that
\begin{equation}\label{gauss lik}
{\cal L}(\{d_j\}\vert\{\theta_i\},{\cal M})\propto e^{-\chi^2(\{\theta_i\})/2}.
\end{equation}

%%%%%%%%%%%%%%%%%%%%%%%%%%%%%%%
%Prior, posterior
%%%%%%%%%%%%%%%%%%%%%%%%%%%%%%%

The probability density function $p(\{\theta_i\}\vert\{d_j\}, {\cal M})$  of the parameters to have values
$\{\theta_i\}$  under the assumption that the true  model is ${\cal M}$  and provided that the available observational data are $\{d_j\}$ reads  \cite{jussi}
\begin{widetext}
\begin{equation}
p(\{\theta_i\}\vert\{d_j\},{\cal M})= \frac{{\cal L}(\{d_j\}\vert\{\theta_i\},{\cal M})\pi(\{\theta_i\},{\cal M})}{\int{\cal L}(\{d_j\}\vert\{\theta_i\},{\cal M})\pi(\{\theta_i\},{\cal M})d\theta_1\dots d\theta_n}\,.
\end{equation}
\end{widetext}
In the Bayesian framework $p(\{\theta_i\}\vert\{d_j\},{\cal M})$ and $\pi(\{\theta_i\},{\cal M})$ are respectively called the posterior and prior probability density functions (pdf) \cite{jussi, trotta1,trotta2, cousins, holmes}. The prior pdf encodes all previous knowledge about the parameters before the observational data have been collected. It may be regarded as subjective up to a certain point, but its use is compulsory in  the Bayesian approach, which is arguably  the approach to be used for theoretical frameworks
which do not admit repetition of experiments (we only have one universe to enquire about). 

Parameter estimation in the Bayesian framework is based on maximizing the posterior pdf $p(\{\theta_i\}\vert\{d_j\},{\cal M})$, whereas in a ``strict''   frequentist approach  one just maximizes $ {\cal L}(\{d_j\}\vert\{\theta_i\},{\cal M})$. When one uses flat priors in the Bayesian approach
then the same conclusions are drawn from both approaches and then the difference turns to be conceptual only  \cite{trotta1,trotta2,trotta3}.  Interestingly, this kind of prior, which is also called the top-hat prior, is the most popular one in usual practise, but this does not necessarily mean it gives a fair representation of  the state of knowledge before the
experiment is carried out, and this usually requires physical insight into
the problem (sometimes priors rest upon symmetry considerations) \cite{trotta3}
\footnote{Some people work in a ``mixed'' approach and consider priors but then follow frequentist procedures.}.

%%%%%%%%%%%%%%%%%%%%%%%%%%%%%%%
%Credible intervals
%%%%%%%%%%%%%%%%%%%%%%%%%%%%%%%
The second step toward constraining parameters satisfactorily is to construct credible intervals \cite{cousins, zech}. We will for a while simplify our notation
and let \begin{equation}p(\{\theta_i\}\vert\{d_j\},{\cal M})\equiv p(\theta_1,\dots,\theta_n).\end{equation}

In the Bayesian approach \cite{cousins} the  $68\%$ credible intervals on the parameter $\theta_i$ will be given as  $\theta_i=x_{-y}^{+z}$ with $x,y,z$ calculated as follows. Here $x$ will be the median of the  marginal probability density function 
\begin{equation}
{p}(\theta_i)=\int {p}(\theta_1,\dots,\theta_n)d\theta_1\dots d\theta_{i-1}d\theta_{i+1}\dots d\theta_{n}\,. 
\end{equation} The median $x$ is calculated from 
\begin{equation}\int_ {\theta_{il}}^{x}{p}(\theta_i)d\theta_i=0.5\times\int{p}(\theta_i)d\theta_i
\,. \end{equation}
Similarly, $y$ is calculated from 
\begin{equation}
\int_{\theta_{il}}^{x-y} p(\theta_i)d\theta_i=((1-0.68)/2)\times\int{ p}(\theta_i)d\theta_i\,, \end{equation} and $z$ is calculated from \begin{equation}\int_{x+z}^{\theta_{iu}}p(\theta_i)d\theta_i=((1-0.68)/2)\times\int{p}(\theta_i)d\theta_i\,. \end{equation}
For marginalized normalized likelihood functions (i.e. posteriors) and likelihood contours see Figs. \ref{fig. lcdm}-\ref{fig. QDGP}.
The parameters $\theta_{il}$ and $\theta_{iu}$ entering the definitions of $y$ and $z$ are respectively a lower and an upper bound which should  either be given by some physical restriction, or alternatively be chosen so that the conditions $p(\theta_{il})\approx 0$ and $p(\theta_{iu})\approx 0$  are satisfied to the desired degree of accuracy. Hence, the probability to get the observed values of the different physical quantities into play is 0.68 if $\theta_i$ lies in the range $[x-y,x+z]$. The modification required so as to calculate
the $95\%$  credible interval is straightforward.

In cases with strongly asymmetric posteriors, one will have to resort to alternative ways to report credible intervals \cite{jussitalk}. This will be the case in two of the models under study in this paper, and details will be given below.

In order to calculate the confidence intervals (correspondent of the credible intervals of Bayesian statistics) in the frequentist approach one has to calculate the mode of ${\cal L}(\theta_i)$, which is the value
of $\theta_i$ which maximizes ${\cal L}(\theta_i)$. We will denote this value as $\theta_{i {\rm md}}$ The boundaries of the $68\%$ confidence interval are calculated by finding the two values of $\theta_i$ for which 
${\cal L}(\theta_i)={\cal L}(\theta_{i {\rm md}}) e^{-1/2}$ \cite{zech},  whereas the $95\%$ confidence interval is obtained by the same recipe, except that the exponent $-1/2$ must be replaced by $-2$. We will not use this way of calculating confidence intervals because it is exact only in gaussian situations (see Eq. (\ref{gauss lik})), but we will rather do it the Bayesian way.\\

In two of our models  (specifically LDGP and QDGP) it happens that, due to nonphysicality reasons, the allowed region of one of the parameters ($\Omega_{r_c}$)  is truncated by a theoretical cut. This sharp restriction demands a modification of the definition of credible intervals.  Let us denote this parameter as $\theta_{\star}$ for simplicity (and 
consistency with the notation in this section). The truncation  is such that $\theta_{\star l}=\theta_{\star  {\rm md}}$, so the
$68\%$ credible interval on $\theta_{\star}$  will be reported as
 $\theta_{\star}=x^{+z}$, where  $x$ will now denote $\theta_{\star  {\rm md}}$
and $z$ will be calculated as 
\begin{equation}
\int_x^z p(\theta_{\star})d\theta_\star=0.68\times\int p(\theta_{\star})d\theta_\star.
\end{equation}\rm 
Table \ref{table2} summarizes results.
%%%%%%%%%%%%%%%%%%%%%%%%%%%%%%%
%\subsection{Model comparison}
%%%%%%%%%%%%%%%%%%%%%%%%%%%%%%% 
\subsection{Model comparison}
A popular but not too refined way to rate goodness of models in the frequentist approach is to compare values of the quantity $-2\log {\rm max}\{{\cal L}(\theta_{1},\dots,\theta_n)\}/{\rm dof}\equiv
{\rm max}\{\chi^2(\theta_{1},\dots,\theta_n)\}$ where dof stands for the number of degrees of freedom of the model, which in turn is the number of observational data point minus the number of parameters. 
Nevertheless, in general ${\rm max}\{{\cal L}(\theta_{1},\dots,\theta_n)\}\neq
{\cal L}(\theta_{1{\rm md}},\dots,\theta_{n{\rm md}})$; if we let
${\rm max}\{{\cal L}(\theta_{1},\dots,\theta_n)\}=
{\cal L}(\theta_{1{\rm bf}},\dots,\theta_{n{\rm bf}})$
then the vector $(\theta_{1{\rm bf}},\dots,\theta_{n{\rm bf}})$ will represent what is usually referred to as the best fit.

In the Bayes approach, the preferred estimator is the evidence \cite{holmes}. 
It does not rely exclusively on the best-fitting parameters of the model, in contrast, it informs about how well the parameters of the model
fit the data, after doing an averaging over all the parameter values that were theoretically plausible before the measurement ever took place 
\cite{liddle}.

%%%%%%%%%%%%%%%%%%%%%%%%%%%%%%%
%Bayes evidence
%%%%%%%%%%%%%%%%%%%%%%%%%%%%%%%
We denote Bayes evidence as ${\cal E({\cal M})}= p(\{d_j\}\vert {\cal M})$, and it is defined as the probability of the data  $\{d_j\}$ given the model
${\cal M}$, that is, 
\begin{equation}
{\cal E(M)}=\int\pi(\{\theta_i\},{\cal M}){\cal L}(\{d_j\}\vert\{\theta_i\},{\cal M}) d\theta_1\dots d\theta_n\,
\end{equation}
where $\pi(\{\theta_i\},{\cal M})$ is the model's prior on the set of parameters, normalized to unity (i.e.
($\int \pi(\{\theta_i\},{\cal M})d\theta_1\dots d\theta_n\!\!=1\!$.)
The most common choice is the top hat prior, so that one rewrites  Bayes evidence as
\begin{equation}
{\cal E(M)}=\frac{\int_{\theta_{1 {\rm min}}}^{\theta_{1 {\rm max}}}\dots \int_{\theta_{n {\rm min}}}^{\theta_{n {\rm max}}} {\cal L}(\theta_1,\dots,\theta_n)d\theta_1\dots \theta_n }{\int_{\theta_{1 {\rm min}}}^{\theta_{1 {\rm max}}}\dots \int_{\theta_{n {\rm min}}}^{\theta_{n {\rm max}}} d\theta_1\dots d\theta_n}.
\end{equation}

Now, to realize of the usefulness of the Bayes evidence towards our final goal, which  is  model selection, note that in Bayesian statistics the preference of model ${\cal M}_i$ over model ${\cal M}_j$ given the data $\{d_k\}$ is estimated through the quotient
\begin{equation}
\frac{p({\cal M}_i\vert\{d_k\})}{p({\cal M}_j\vert\{d_k\})}=\frac{{\cal E}_i ({\cal M}_i)}{{\cal E}_j({\cal M}_j)}\frac{\pi_i({\cal M}_i)}{\pi_j({\cal M}_j)}.
\end{equation}
On the other hand, the Bayes factor $B_{ij}$ for any two models ${\cal M}_i$ and ${\cal M}_j$ is defined as 
\begin{equation}
B_{ij}=\frac{{\cal E}_i ({\cal M}_i)}{{\cal E}_j ({\cal M}_j)}
\end{equation}
so if, as in usual practise, one assumes no prior preference of one model over the other, that is, $\pi_i({\cal M}_i)=\pi_j({\cal M}_j)=1/2$, one finally has
\begin{equation}
\frac{p({\cal M}_i\vert\{d_k\})}{p({\cal M}_j\vert\{d_k\})}=B_{ij}.
\end{equation}

We report values of the evidences and Bayes factor for the models considered here in Table \ref{evidences} (see Table \ref{priors} for priors).
Values of the Bayes factor could be used to attribute evidence to the model ${\cal M}_i$ against the model ${\cal M}_j$ using Jeffreys' scale \cite{Jeff} which 
regards evidences of one model against the other as not significant, substantial, strong or decisive depending on values of $ln(B_{ij})$. %
%can be summarized as follows:
%\begin{itemize}
%\item If $ln(B_{ij})<1$, then the evidence against ${\cal M}_j$ is not significant.
%\item If $1<ln(B_{ij})<2.5$, then the evidence against ${\cal M}_j$ is substantial.
%\item If $2.5<ln(B_{ij})<5$, then the evidence in favor of ${\cal M}_i$ is strong.
%\item If $5<ln(B_{ij})$, then the evidence in favor of ${\cal M}_i$ is decisive.
%\end{itemize}
Recently a the refinement of this scale has been  proposed by Wasserman \cite{Wass}. These scales are of course subjective and one can question how illustrating they actually are, but the underlying philosophy is that the Bayes factor is a ratio of odds in the case there is no preference of one model over the other, so for the benefit of the readers which prefer a more objective report of our results we  give, in Table \ref{table5}, the ratio of odds of different pairs of models among those under consideration, which we will express as ${\cal O}_{ij}\equiv B_{ij}:1$ in the cases where $B_{ij}\ge1$, whereas for the sake of aesthetics we will use the alternative look ${\cal O}_{ij}\equiv 1:B_{ji}$. These expressions then must be interpreted in the sense that  an odds ratio equivalent to $B_{ij}:1$ means the model  $i$   is $B_{ij}$ times more probable than model $j$, whereas and odds ratio of the form $1:B_{ji}$ means model $i$ is  $B_{ji}$ times less probable than model $j$.
%%%%
%Our Bayesian evidence limits
% 0<om<0.6
%0<ol<2
%-2<w<0
%-2<w0<0
%2<w1<4

%%%%%%%%%%%%%%%%%%%%%%%%%%%%%%%
%Cases with more parameters
%%%%%%%%%%%%%%%%%%%%%%%%%%%%%%%

Before we close this section, an important remark is in order. The usual situation in cosmology is that for the task of constraining the parameters $\{\theta_i\}$ one has at hand more than one set of statistically independent observational data,  say $\{d^{(1)}_j\},\dots \{d^{(m)}_k\}$; in that case, one can resort to the
joint probability density function 
\begin{eqnarray}
p(\{\theta_i\}\vert\{d^{(1)}_j\}\cap\dots\cap\{d^{(m)}_k\}, {\cal M})=\nonumber\\p(\{\theta_i\}\vert\{d^{(1)}_j\}, {\cal M})\times
\dots\times
p(\{\theta_i\}\vert\{d^{(m)}_k\}, {\cal M}).
\end{eqnarray}
Clearly, the latter rule can be used to generalize conveniently the whole discussion   above  to this situation with
availability of more than one dataset.

%%%%%%%%%%%%%%%%%%%%%%%%%%%%%%%
\section{Results and Conclusions}\label{rescon} 
%%%%%%%%%%%%%%%%%%%%%%%%%%%%%%%

 \begin{widetext}

\begin{table}[htp!]\caption{Best fits, $\chi^2$ and Credible Intervals}\label{table2}
	\centering
		\begin{tabular}{l@{\hspace{0.4cm}}c@{\hspace{0.4cm}}c@{\hspace{0.4cm}}c@{\hspace{0.4cm}}c}
		\hline\hline
Model\quad& Best-fit     & $\chi^2/{\rm dof}$& $68\%$ conf. int.     & $95\%$ conf. int.\\
\hline\\
LCDM & $\Omega_m=0.26$   &  $1.03$  &$\Omega_m=0.26_{-0.02}^{+0.02}$ &$\Omega_m=0.26_{-0.04}^{+0.05}$\\\\
%%%%%%%%%%%%%%%%%%%%%%%%%%%%%
DGP  & $\Omega_m=0.23$   & $1.08$   &$\Omega_m=0.23_{-0.02}^{+0.02}$ &$\Omega_m=0.23_{-0.04}^{+0.04}$\\\\
%%%%%%%%%%%%%%%%%%%%%%%%%%%%%
QCDM & $\Omega_m=0.24$   & $1.03$   &$\Omega_m=0.25_{-0.02}^{+0.03}$ &$\Omega_m=0.25_{-0.05}^{+0.06}$\\
     & $w= -0.94$        &          &$w=-0.96_{-0.09}^{+0.08}$       &$w=-0.96_{-0.20}^{+0.15}$      \\\\
%%%%%%%%%%%%%%%%%%%%%%%%%%%%%
LDGP & $\Omega_m=0.26$   & $1.04$   &$\Omega_m=0.26_{-0.02}^{+0.03}$ &$\Omega_m=0.26^{+0.05}_{-0.05}$\\
     & $\Omega_{r_c}=0.00$&          &$\Omega_{r_c}=0.00^{+0.02}$      &$\Omega_{r_c}=0.00^{+0.05}$     \\\\
%%%%%%%%%%%%%%%%%%%%%%%%%%%%%
QDGP & $\Omega_m=0.25$   & $1.04$   &$\Omega_m=0.23_{-0.03}^{+0.03}$ &$\Omega_m=0.23_{-0.05}^{+0.06}$\\ 
     & $\Omega_{r_c}=0.00$&          &$\Omega_{r_c}=0.00^{+0.88}$   &$\Omega_{r_c}=0.00^{+2.32}$     \\
     & $w=-0.95$          &          &$w=-0.81_{-0.07}^{+0.05}$       &$w=-0.81_{-0.19}^{+0.09}$      \\\\
%%%%%%%%%%%%%%%%%%%%%%%%%%%%%
CPL  & $\Omega_m=0.25$   & $1.04$   &$\Omega_m=0.26_{-0.03}^{+0.03}$ &$\Omega_m=0.26_{-0.05}^{+0.06}$\\ 
     & $w_0=-1.08$       &          &$w_0=-1.05_{-0.20}^{+0.19}$     &$w_0=-1.05_{-0.47}^{+0.37}$    \\
     & $w_1=0.58$        &          &$w_1=0.37_{-0.84}^{+0.70}$      &$w_1=0.37_{-1.78}^{+1.34}$     \\\\
\\
\hline
		\end{tabular}
\end{table}

\begin{table}[htp!]\caption{Priors}\label{priors}
\begin{center}
\begin{tabular}{l@{\hspace{0.2in}}c}
\hline
 Model &   Parameter ranges \\
\hline
LCDM   & $0. 00 \le  \Omega_m \le0.60$\\
DGP   &  $0. 00 \le  \Omega_m \le0.60$\\
QCDM   & $0.00\le  \Omega_m \le0.60$ , \; $-1.50\le w\le -0.33$  \\
LDGP   &  $0.00 \le  \Omega_m \le0.60$, \; $0.00\le\Omega_{r_c}\le 0.10$  \\
CPL    &  $0.00\le  \Omega_m \le0.60$,\;  $-1.50\le w_0\le -0.33$, \;     $0.00\le w_1\le 2.00$ \\
QDGP   &  $0.00\le  \Omega_m \le0.60$, \; $-1.50\le w\le -0.33$, \; $0.00\le\Omega_{r_c}\le 1.00$  \\
\hline
\end{tabular}
\end{center}
\end{table}

\begin{table}[htp!]\caption{Bayes evidence}
	\centering
		\begin{tabular}{l@{\hspace{0.2in}}c@{\hspace{0.2in}}c@{\hspace{0.2in}}c@{\hspace{0.2in}}c@{\hspace{0.2in}}c@{\hspace{0.2in}}c}
		\hline
Model  & ${\cal E}$           &$ln(B_{i,{\rm DGP}})$  &$ln(B_{i,{\rm QCDM}})$  &$ln(B_{i,{\rm LDGP}})$  &$ln(B_{i,{\rm CPL}})$  &$ln(B_{i,{\rm QDGP}})$  \\
\hline
LCDM & $1.05\times10^{-46}$ &  5.55           &  1.54            &  2.31            &  1.83           &   3.56           \\
DGP  & $4.07\times10^{-49}$ &     0.00            & -4.01            & -3.24            & -3.73           & -2.00        \\
QCDM & $2.25\times10^{-47}$ &                 &       0.00           &  0.77            &  0.29           & 2.01            \\
LDGP &$1.04\times10^{-47}$  &                 &                  &         0.00         & -0.49           & 1.24            \\
CPL  &$1.69\times10^{-47}$  &                 &                  &                  &       0.00          & 1.72         \\
QDGP &$3.02\times10^{-48}$  &                 &                  &                  &                 &    0.00              \\
\hline
		\end{tabular}
\label{evidences}		
 \end{table}

\begin{table}[htp!]\caption{Odds ratio (number of times a model is more probable than other)}\label{table5}
	\centering
		\begin{tabular}{l@{\hspace{0.2in}}c@{\hspace{0.2in}}c@{\hspace{0.2in}}c@{\hspace{0.2in}}c@{\hspace{0.2in}}c}
		\hline
Model           &${\cal O}_{i,{\rm DGP}}$  &${\cal O}_{i,{\rm QCDM}}$  &${\cal O}_{i,{\rm LDGP}}$  &${\cal O}_{i,{\rm CPL}}$  &${\cal O}_{i,{\rm QDGP}}$  \\
\hline
LCDM   &  258:1            &  5:1            &  10:1    &6:1      &   35:1           \\
DGP   &     1:1            &  1:55          &       1:26    &       1:42    &   1:7          \\
QCDM  &                 &      1:1            &        2:1      &      1:1     & 7:1             \\
LDGP  &                &                  &     1:1             &     1:2      &    3:1         \\
CPL    &                 &                  &                 &    1:1            & 6:1            \\
QDGP   &                &                  &                  &                 &  1:1                \\
\hline
		\end{tabular}
 \end{table}
\label{odds}
 \end{widetext}

In this paper we have performed parameter estimation and model selection
for a set of flat FRW cosmological models which could be suitable to explain the observed current acceleration. Half of the models stem from the assumption of a dark energy component in the Universe, whereas the other half is made by models motivated by extra dimensions and that include a gravitational component of the observed expansion acceleration. We analyze data coming from (most recent) SNIa luminosity, CMB shift and $H(z)$ measurements. Even though most of our statistical treatment fits  into the Bayesian  approach, for completeness we also report results on some frequentist estimators. Our results are consistent with partially overlapping work (e.g. \cite{similar_papers, dgp_goobar}) done previously using different combinations of observations.

One may think a priori the dominant source of error to be the $H(z)$ data, as the percentage errors range from
$8\%$ to $20\%$, whereas for the SN data the lowest percentage error is $0.3\%$ and the highest is
$1\%$ on the one hand, and for the CMB shift single datum we have a percentage error of $1.7\%$. On the other hand, since the $\chi^2$ construction from the $H(z)$  data does not involve any integration,
there might be a restraint in the propagation of errors coming from them. We are inclined to think
this is actually the case as the size of errors  in our study are  comparable with those obtained in other works.

In the light of our results we conclude, first of all, that the frequentist estimator $\chi^2/dof$ significantly disfavours the DGP model over all the other models considered. We compute the probability of getting values of the $\chi^2$ not farther from the minimum than the minimum $\chi^2$ for all other models, as defined in \cite{perivol} and we get a probability $P \geq 99\% $. Consistently we arrive at the same conclusion using
Bayes evidence. On the other hand, if we compare LCDM with all the other models, it turns out that from the frequentist perspective it is as good as QCDM, whereas for LDGP, CPL and QDGP  one gets exactly the same value 
$\chi^2/dof$, a value which turns out to be just slightly larger than 
the one for LCDM. In conclusion, the frequentist analysis yields the result that LCDM is just marginally preferred over all the rest of the models, except for DGP which is clearly the least likely model of all six.

Regarding the  Bayesian analysis, the reader can decide whether to qualify our results using the adjectives in Jeffrey's scale or just stick to the objectivity of the numbers
giving us the odds of the different models against the others, but at least in the  case of DGP versus LCDM we dare to say it is difficult not to be carried by the impression
that it is rather disfavored.

Finally, our results on credible limits point us in the direction that this combination of data sets constrain very poorly  models with two ``dark energy'' parameters, so parameter constraints on 
CPL and QDGP seem not to be tight enough.

\section*{Acknowledgements}
We are grateful to Roy  Maartens for reading
the manuscript, engaging in useful conversations and making encouraging comments. We acknowledge Luca Amendola for offering us his software from which part of our codes originates.  Thanks to Mariam Bouhmadi-L\'opez, Rob Crittenden, Bob Nichol, Will Percival, Bj\"orn Sch\"afer, Roberto Trotta and Jussi V\"aliviita  for enlightening discussions. 

R.L. is supported
by the University of the Basque Country through
research grant GIU06/37  and by the
Spanish Ministry of Education and Culture through the
RyC program, and research grants FIS2004-01626  and
FIS2004-0374-E.
 \begin{widetext}
 \appendix
 \section{Probability densities}
 In this appendix we present the plots of the probability densities of the six different models considered.  As explained above ${\cal L}_{\theta_i}$ will represent the probability density obtained by marginalization over all the parameters of the model except for the parameter $\theta_i$. For the models with more than one parameter we also plot the likelihood contours associated to 68, 95 and 99$\%$ probability. For the models with 3 parameters we have fixed $\Omega_m$ to be $0.25$.

%Our notation is summarized in the following table
% \begin{table}[hb]\caption{Denomination of the parameters for each model}
%	\centering
%		\begin{tabular}{c@{\hspace{0.2in}}c@{\hspace{0.2in}}c@{\hspace{0.2in}}c}
%		\hline
%Model& $\theta_1$ & $\theta_2$& $p_3$\\
%\hline
%LCDM & $\Omega_m$&&\\
%DGP & $\Omega_m$&&\\
%QCDM & $\Omega_m$ & $w$ &  \\
%LDGP& $\Omega_m$&$\Omega_{r_c}$\\
%CPL & $\Omega_m$  &$w_0$ & $w_1$ \\
%QDGP & $\Omega_m$ &$\Omega_{rc}$&  $w$ \\
%\hline
%		\end{tabular}
% \end{table}

%%%%%%%%%%%%%%%%%%%%%%%%%%%
\begin{figure*}[t!]
\begin{center}
\includegraphics[width=7cm]{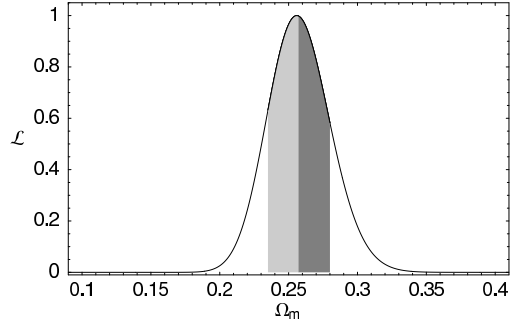}\quad
\caption{Likelihood function for the flat LCDM model. The light (dark) shaded region indicates the error on $\Omega_m$ to the left (right) of the median.
}\label{fig. lcdm}
\end{center}
\end{figure*}
%%%%%%%%%%%%%%%%%%%%%%%%%%%

%%%%%%%%%%%%%%%%%%%%%%%%%%%
\begin{figure*}[t!]
\begin{center}
\includegraphics[width=7cm]{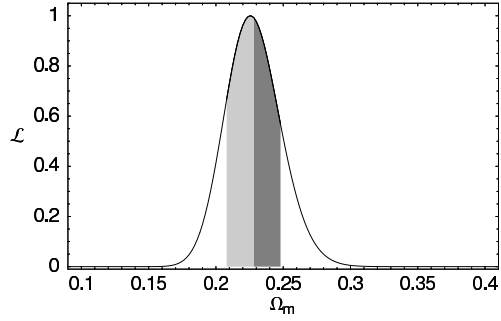}\quad
\caption{Same as for fig.\ref{fig. lcdm} but for the flat DGP model.
}\label{fig. dgp}
\end{center}
\end{figure*}
%%%%%%%%%%%%%%%%%%%%%%%%%%%

%%%%%%%%%%%%%%%%%%%%%%%%%%%
\begin{figure*}[t!]
\begin{center}
\includegraphics[width=7cm]{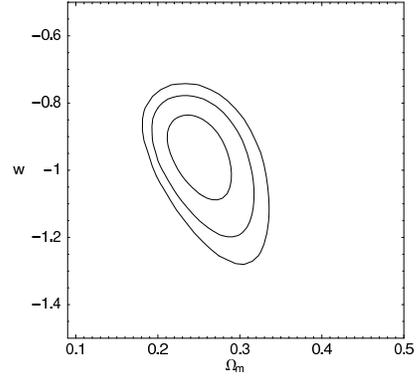}\quad
\caption{Likelihood contours corresponding to the 68, 95 and 99$\%$ credible contours in the ($w$, $\Omega_m$) plane for the QCDM model.
}\label{fig.contp_quies}
\end{center}
\end{figure*}
%%%%%%%%%%%%%%%%%%%%%%%%%%%

%%%%%%%%%%%%%%%%%%%%%%%%%%%
\begin{figure*}[t!]
\begin{center}
\includegraphics[width=7cm]{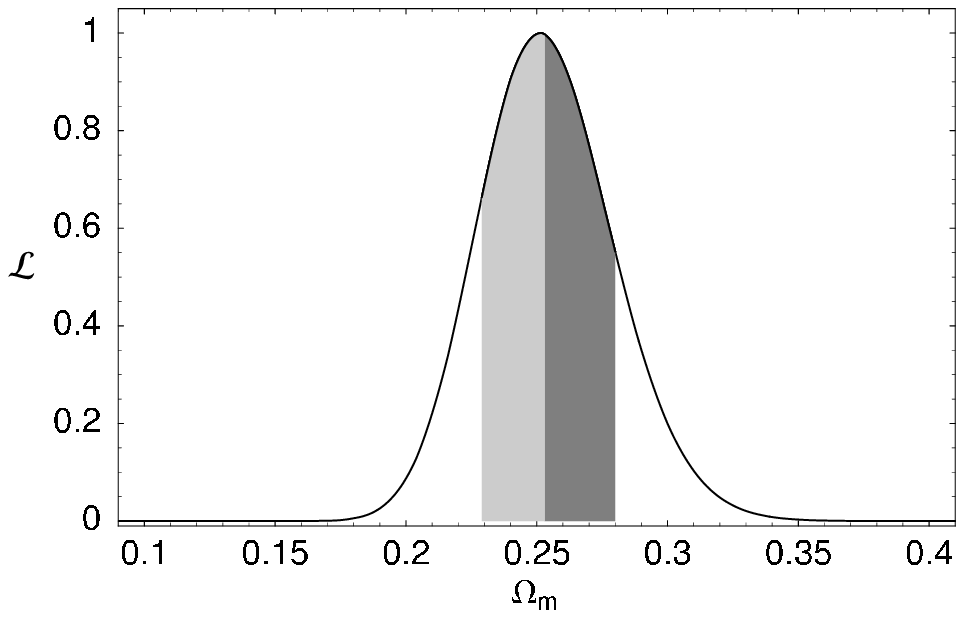}\quad
\includegraphics[width=7cm]{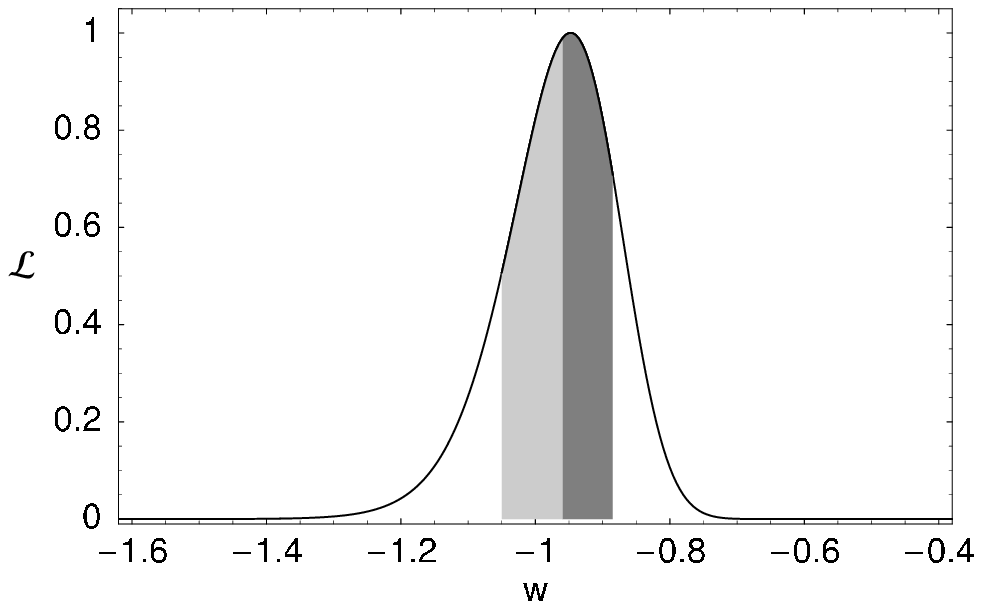}\quad \\
\caption{Marginalized likelihood functions for the QCDM model for  $\Omega_m$ (left) and the equation of state $w$ (right).
}\label{fig. QCDM}
\end{center}
\end{figure*}
%%%%%%%%%%%%%%%%%%%%%%%%%%%

%%%%%%%%%%%%%%%%%%%%%%%%%%%
\begin{figure*}[t!]
\begin{center}
\includegraphics[width=7cm]{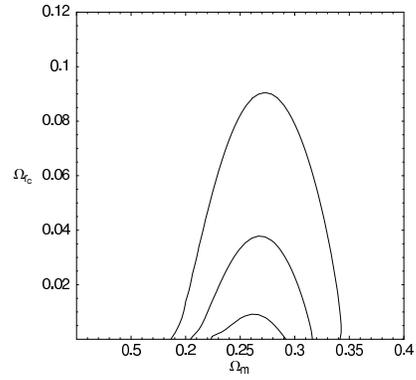}\quad
\caption{Likelihood contours  corresponding to the 68, 95 and 99$\%$ credible contours in the ($\Omega_m$, $\Omega_{r_c}$) plane for the LDGP model.}\label{fig.contp_ldgp}
\end{center}
\end{figure*}
%%%%%%%%%%%%%%%%%%%%%%%%%%%

%%%%%%%%%%%%%%%%%%%%%%%%%%%
\begin{figure*}[t!]
\begin{center}
\includegraphics[width=7cm]{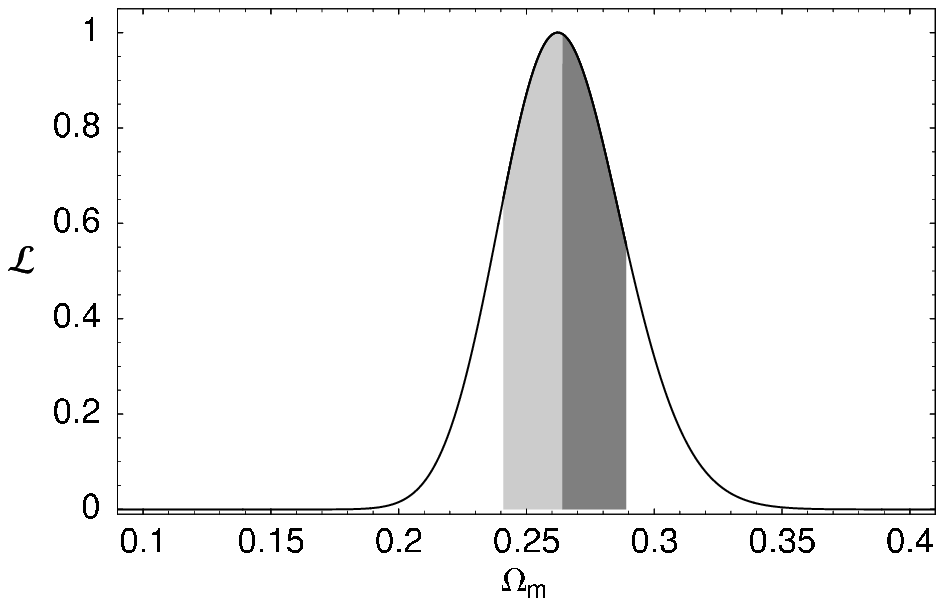}\quad
\includegraphics[width=7cm]{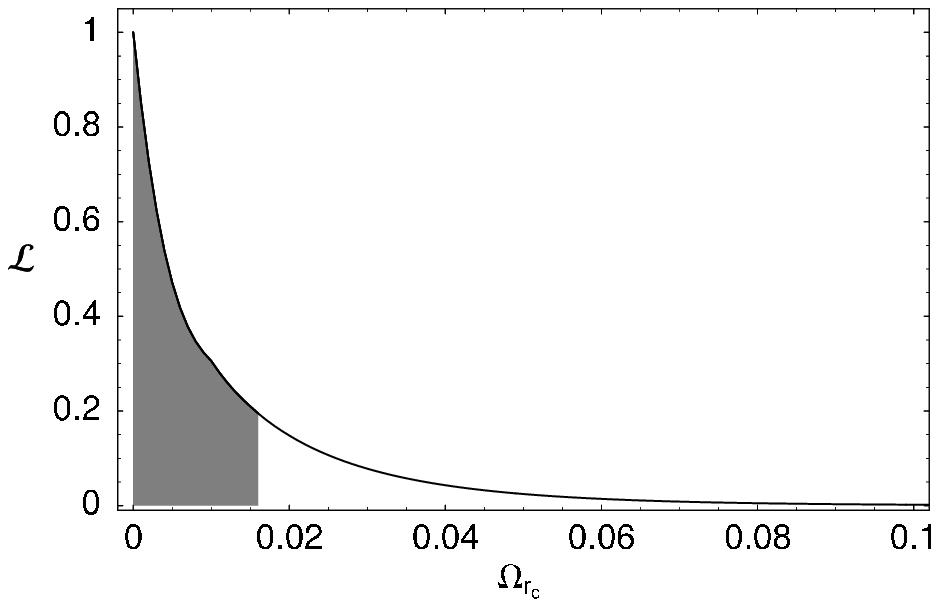}\quad \\
\caption{Marginalized likelihood functions for the LDGP model for $\Omega_m$ (left) and $\Omega_r$ (right).
}\label{fig. LDGP}
\end{center}
\end{figure*}
%%%%%%%%%%%%%%%%%%%%%%%%%%%

%%%%%%%%%%%%%%%%%%%%%%%%%%%
\begin{figure*}[t!]
\begin{center}
\includegraphics[width=7cm]{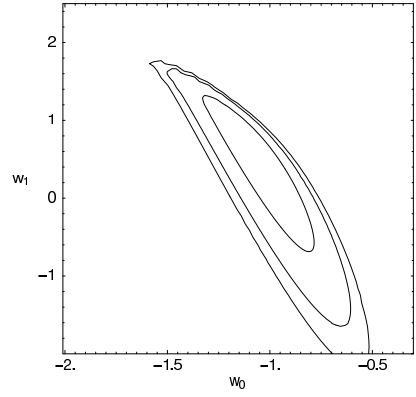}\quad
\caption{Likelihood contours corresponding to the 68, 95 and 99$\%$ credible contours in the ($w_0$, $w_1$) plane for the CPL Ansatz. $\Omega_m$ has been fixed to $0.25$.
}\label{fig.contp_linder}
\end{center}
\end{figure*}
%%%%%%%%%%%%%%%%%%%%%%%%%%%

%%%%%%%%%%%%%%%%%%%%%%%%%%%
\begin{figure*}[t!]
\begin{center}
\includegraphics[width=7cm]{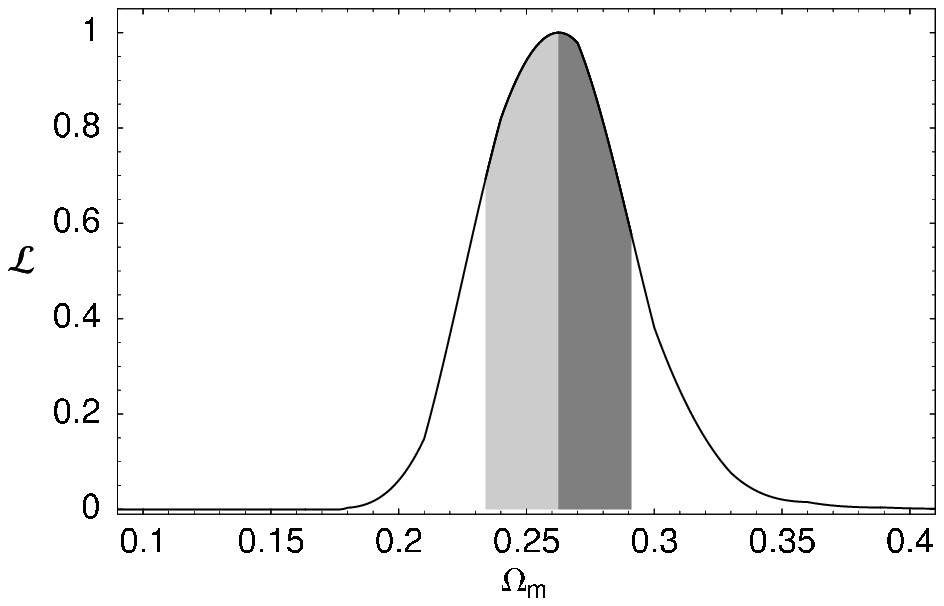}\quad
\includegraphics[width=7cm]{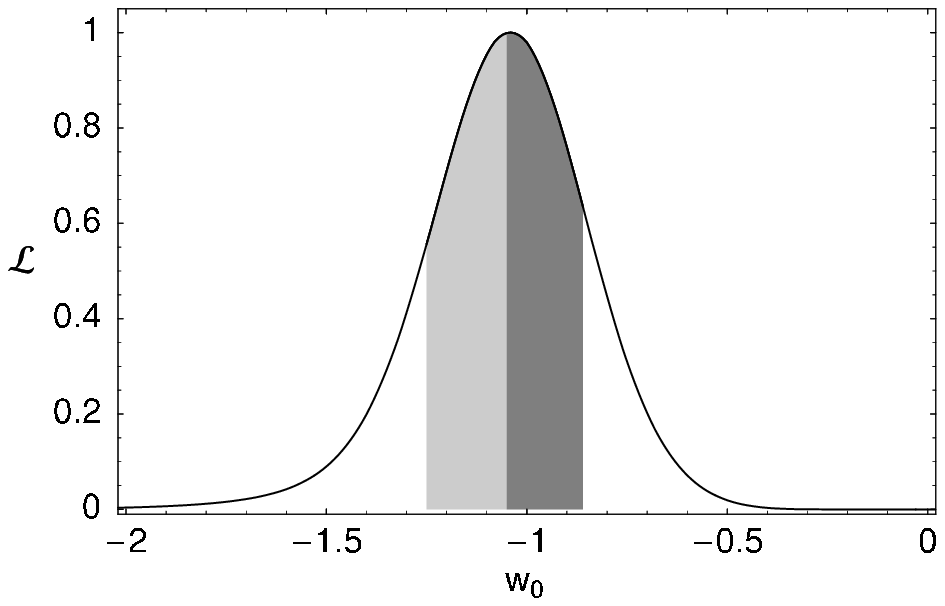}\quad \\
\includegraphics[width=7cm]{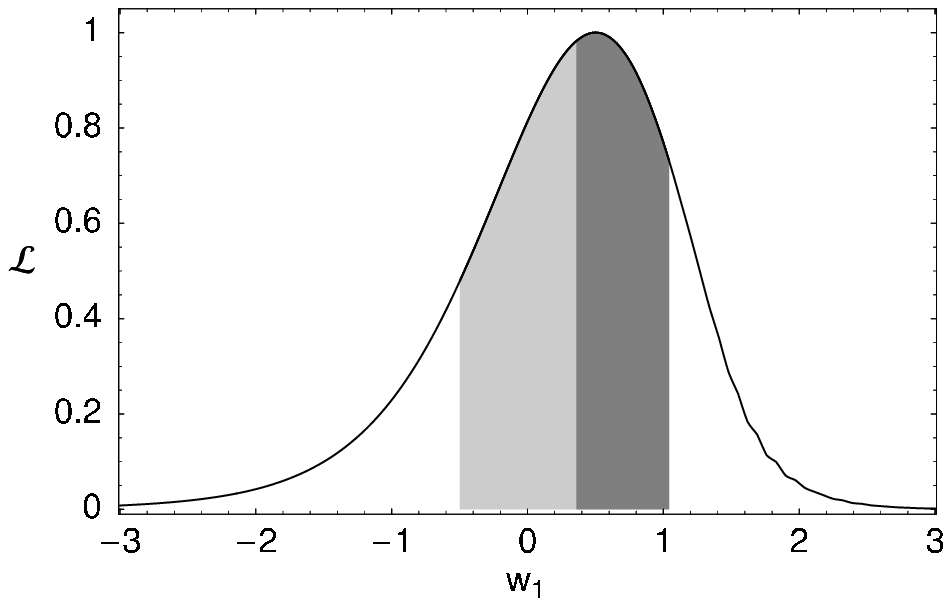}
\caption{Marginalized likelihood functions for the CPL model for $\Omega_m$ (left top), $w_0$ (right top) and  $w_1$ (bottom).
}\label{fig. LCP}
\end{center}
\end{figure*}
%%%%%%%%%%%%%%%%%%%%%%%%%%%

%%%%%%%%%%%%%%%%%%%%%%%%%%%
\begin{figure*}[t!]
\begin{center}
\includegraphics[width=7cm]{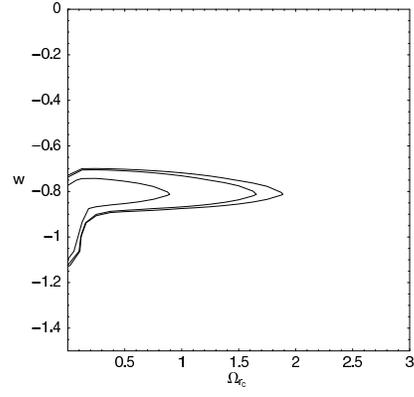}\quad
\caption{Likelihood contours corresponding to the 68, 95 and 99$\%$ credible contours in the ($\Omega_{r_c}$, $w$) plane for the QDGP model. $\Omega_m$ has been fixed to $0.25$.
}\label{fig.contp_qdgp}
\end{center}
\end{figure*}
%%%%%%%%%%%%%%%%%%%%%%%%%%%

%%%%%%%%%%%%%%%%%%%%%%%%%%%
\begin{figure*}[t!]
\begin{center}
\includegraphics[width=7cm]{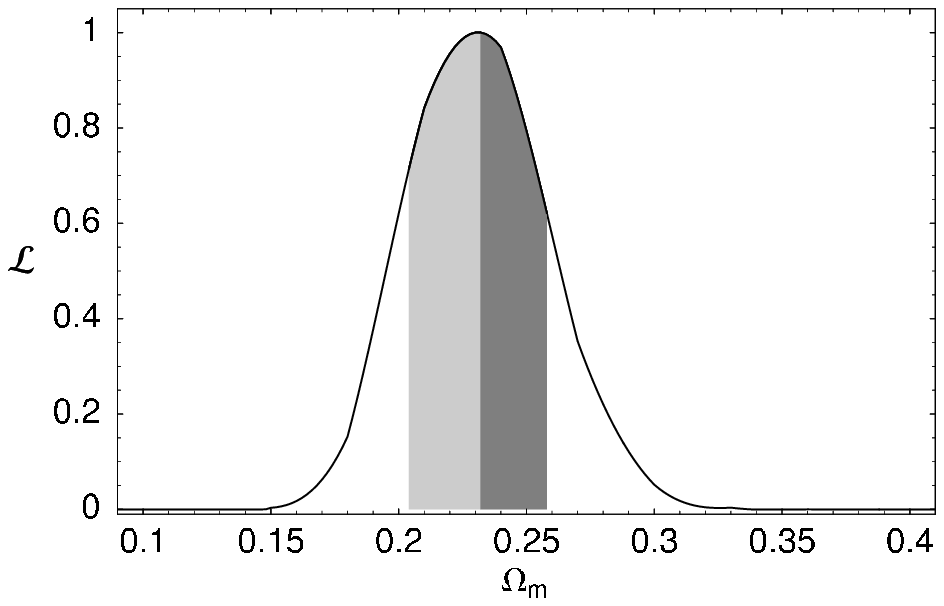}\quad
\includegraphics[width=7cm]{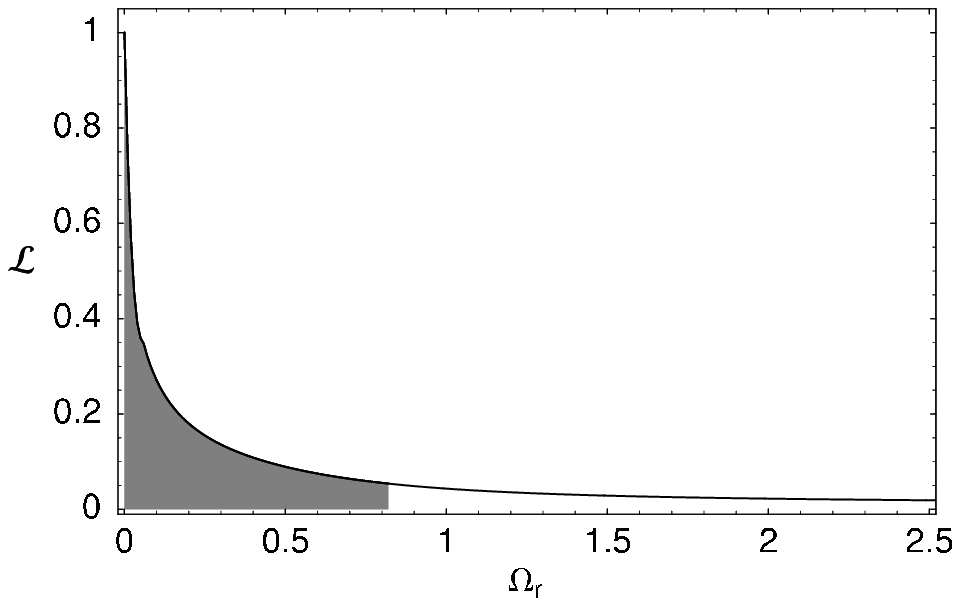}\quad \\
\includegraphics[width=7cm]{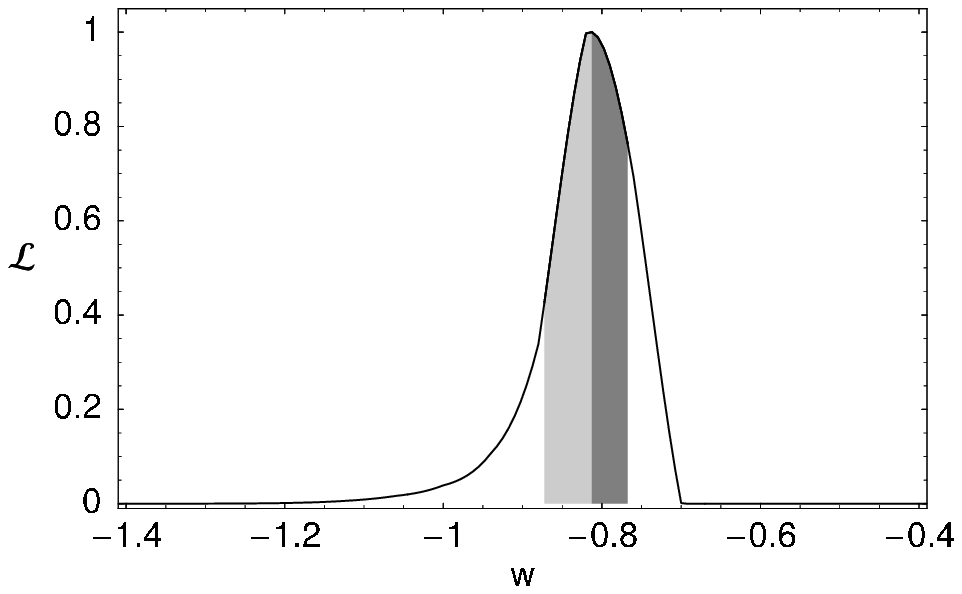}
\caption{Marginalized likelihood functions for the QDGP model for $\Omega_m$ (left top), $\Omega_r$ (right top) and $w$ (bottom).
}\label{fig. QDGP}
\end{center}
\end{figure*}
%%%%%%%%%%%%%%%%%%%%%%%%%%%

\clearpage

\section{Equation of state parameters}

For the sake of illustration, we represent the best fits $w_{de}(z)$ for the CPL model and 
$w_{\rm eff}(z)$ for the DGP and LDGP models. We do not present the plots for the LCDM and QCDM models as the corresponding plots are basically stripes of constant width as they correspond to non-dynamical dark energy. For every best fit we include in the plot an estimation of error propagation based on the assumption of no error correlation.  As in our parameter constraints, we have various sources of complication. On the one hand, as in some models (CPL, for instance) we have found  slight departures from gaussianity, 
we cannot use the standard error propagation formula, in addition the errors found are large in the case of some parameters
also we put forward a modification of  the standard formula in order to account for non-gaussianities and large errors. Assuming our constraints in a generic cosmological parameter $\theta_i$  are reported in the form ${\theta_{i{\rm md}}}_{-\Delta\theta_{il}}^{+\Delta\theta_{iu}}$, with $\Delta\theta_{il}$ and $\Delta\theta_{iu}$
being positive quantities. We can give an estimate of the error in a quantity $f(\theta_1,\dots,\theta_n)$ derived from various such parameters by taking for the upper error the quantity
\begin{equation}
\Delta f_u= \sqrt{\sum_{i=1}^n\left({\rm max}\left(\Delta f_{iu},-\Delta f_{il}\right)\right)^2}
\end{equation}
and for the lower error the quantity 
\begin{equation}
\Delta f_l= \sqrt{\sum_{i=1}^n\left({\rm min}\left(\Delta f_{iu},-\Delta f_{il}\right)\right)^2},
\end{equation}
where
\begin{equation}
 \Delta f_{iu}=f(\dots\theta_{(i-1){\rm md}},\theta_{i{\rm md}}+\Delta\theta_{iu},
\theta_{(i+1){\rm md}},\dots)-f(\dots\theta_{(i-1){\rm md}},\theta_{i{\rm md}},
\theta_{(i+1){\rm md}},\dots)
\end{equation}
and
\begin{equation}
 \Delta f_{il}=f(\dots\theta_{(i-1){\rm md}},\theta_{i{\rm md}}-\Delta\theta_{il},
\theta_{(i+1){\rm md}},\dots)-f(\dots\theta_{(i-1){\rm md}},\theta_{i{\rm md}},
\theta_{(i+1){\rm md}},\dots).
\end{equation}

So, we have proposed  an estimation of error based on finite differences; nevertheless, one can be more refined if errors are small, i.e. $\Delta\theta_{iu}=\delta\theta_{iu}$ and $\Delta\theta_{il}=\delta\theta_{il}$, because then one can write
\begin{equation}
\Delta f_u\simeq\delta f_u= \sqrt{\sum_{i=1}^n\left({\rm max}\left(\frac{\partial f}{\partial \theta_i}\delta\theta_{iu},-\frac{\partial f}{\partial \theta_i}\delta\theta_{il}
\right)\right)^2}.
\end{equation}
and 
\begin{equation}
\Delta f_l\simeq\delta f_l= \sqrt{\sum_{i=1}^n\left({\rm min}\left(\frac{\partial f}{\partial \theta_i}\delta\theta_{iu},-\frac{\partial f}{\partial \theta_i}\delta\theta_{il}
\right)\right)^2}.
\end{equation}

Finally, in the gaussian situations  ($\Delta\theta_{iu}=\Delta\theta_{il}=\Delta\theta_{i}$)  (as for instance DGP) we get $\Delta f_u=\Delta f_l$.

In Fig. (\ref{wplots}) we depict the best fit equation of  state parameter along with error stripes for the models with more than one parameter. A few points are worth mentioning regarding the LDGP and QDGP models so as to clarify the peculiarity of the result. First of all, one must keep in mind for both cases we have found the best fit to correspond to their LCDM limit, which translates into an exactly constant equation of state parameter. On the other hand, the LDGP may look even stranger, as no upper error stripe appears in its plot. The reason for this is well grounded analytically, as the LDGP model does not have the ability to cross the phantom divide 
\cite{qdgp paper}, so $w_{\rm eff}$ cannot be error-propagated across that barrier. Our plot just gives a graphical account of that analytical imposibility.

%%%%%%%%%%%%%%%%%%%%%%%%%%%
\begin{figure*}[htp!]
\begin{center}
\includegraphics[width=7cm]{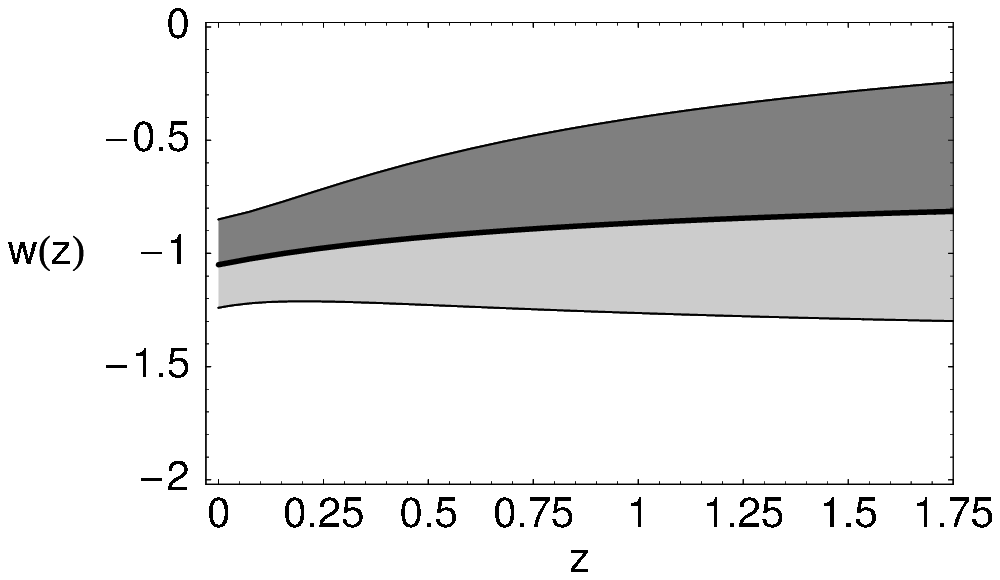}\quad\\
\includegraphics[width=7cm]{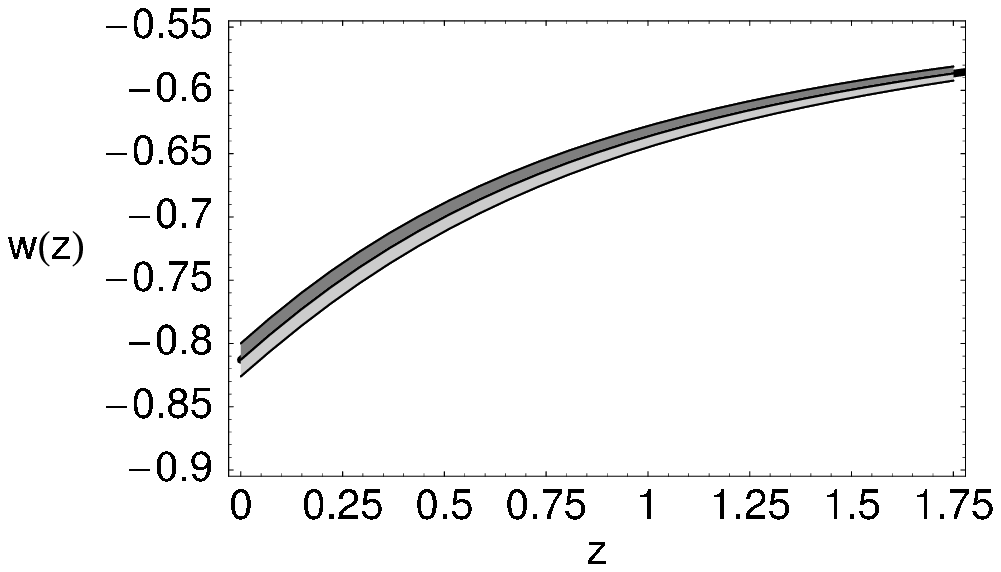}\quad \\
\includegraphics[width=7cm]{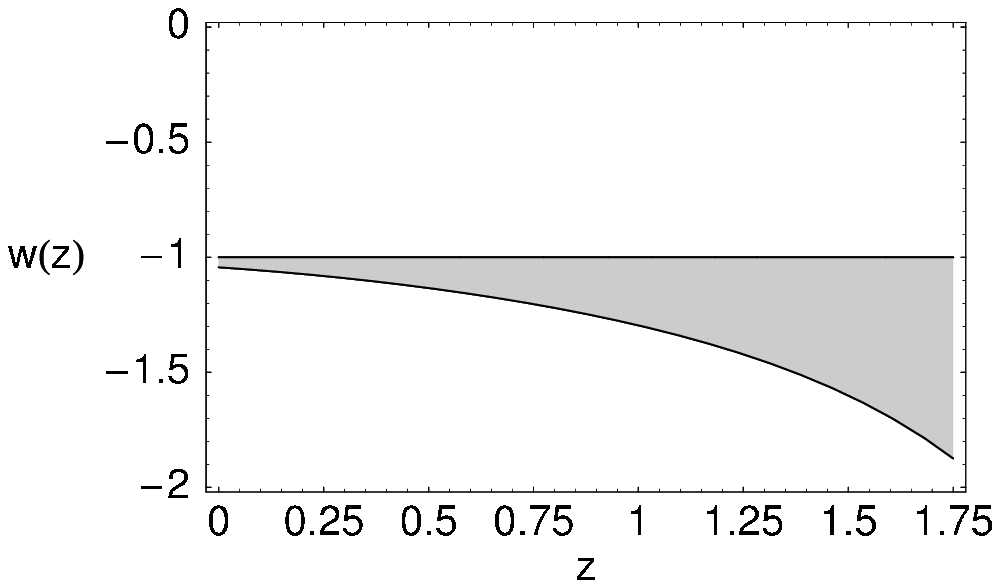}\quad\\
\includegraphics[width=7cm]{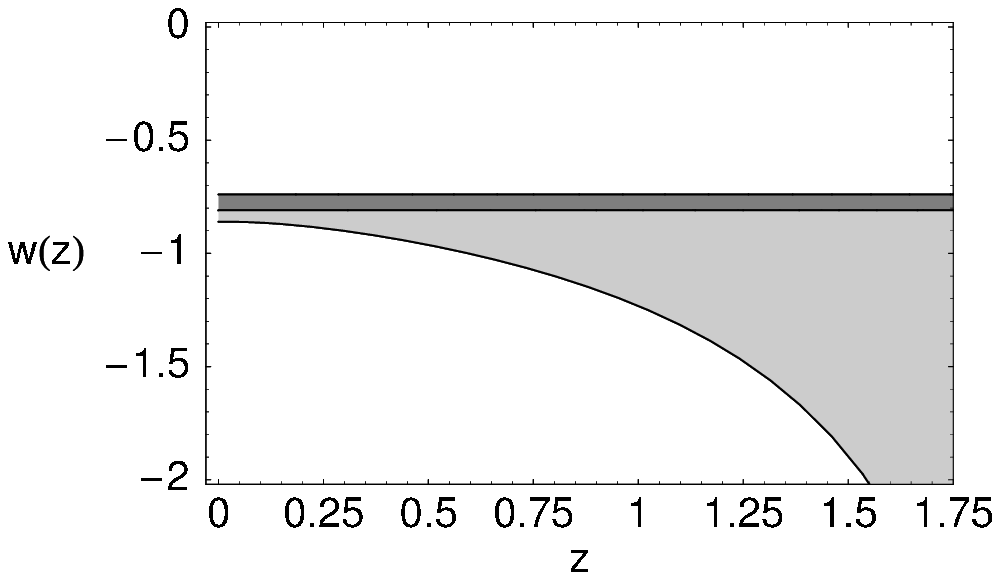}\quad\\
\caption{Equation of state parameter as a function of redshift  for models CPL, DGP, LDGP and QDGP (top to bottom). The black lines correspond to the best fit for each of them, and the gray bands indicate the errors (calculated by the finite differences method). 
}\label{wplots}
\end{center}
\end{figure*}

\end{widetext} 
%%%%%%%%%%%%%%%%%%%%%%%%%%%%%%%%%%%%%%%%%%%%%%%%%%%%%%%%%%%%%%%%%%%%%%%%%%%%%%%%%%%%%%%%%%%%%%%%%%%%%%%%%%%%%%%%%


\begin{thebibliography}{}


\bibitem{simon}
  J.~Simon, L.~Verde and R.~Jim\'enez,
  %``Constraints on the redshift dependence of the dark energy potential,''
  Phys.\ Rev.\ D {\bf 71} (2005) 123001 
  [arXiv:astro-ph/0412269].

%1
\bibitem{ratra}
  L.~Samushia and B.~Ratra,
  %``Cosmological Constraints from Hubble Parameter versus Redshift Data,''
  Astrophys.\ J.\  {\bf 650} (2006) L5
  [arXiv:astro-ph/0607301].
\bibitem{use hubble}
 H.~Wei and S.~N.~Zhang,
  %``Observational $H(z)$ Data and Cosmological Models,''
  Phys.\ Lett.\  B {\bf 644}  (2007) 7
  [arXiv:astro-ph/0609597];

  P.~Wu and H.~W.~Yu,
  %``Generalized Chaplygin gas model: Constraints from Hubble parameter versus
  %redshift data,''
  Phys.\ Lett.\  B {\bf 644} (2007) 16
  [arXiv:gr-qc/0612055];
 P.~X.~Wu and H.~W.~Yu,
  %``Constraints on the unified dark energy-dark matter model from latest
  %observational data,''
  JCAP {\bf 0703} (2007) 015
  [arXiv:astro-ph/0701446];
L.~I.~Xu, C.~W.~Zhang, B.~R.~Chang and H.~Y.~Liu,
  %``Reconstruction of Deceleration Parameters from Recent Cosmic
  %Observations,''
  arXiv:astro-ph/0701519;
 A.~Kurek and M.~Szydlowski,
  %``The LambdaCDM model on the lead -- a Bayesian cosmological models
  %comparison,''
  arXiv:astro-ph/0702484;
H.~Zhang and Z.~H.~Zhu,
  %``Natural Phantom Dark Energy, Wiggling Hubble Parameter $H(z)$ and Direct
  %$H(z)$ Data,''
  arXiv:astro-ph/0703245.

\bibitem{Davis}
T.~M.~Davis {\it et al.},
  %``Scrutinizing exotic cosmological models using ESSENCE supernova data
  %combined with other cosmological probes,''
  arXiv:astro-ph/0701510.



\bibitem{Wang}
  Y.~Wang and P.~Mukherjee,
  %``Robust Dark Energy Constraints from Supernovae, Galaxy Clustering, and
  %Three-Year Wilkinson Microwave Anisotropy Probe Observations,''
  Astrophys.\ J.\  {\bf 650} (2006) 1
  [arXiv:astro-ph/0604051].

\bibitem{Spergel}
  D.~N.~Spergel {\it et al.}  [WMAP Collaboration],
  %``Wilkinson Microwave Anisotropy Probe (WMAP) three year results:
  %Implications for cosmology,''
Astrophys.\ J.\ Suppl.\  {\bf 170} (2007) 377
  [arXiv:astro-ph/0603449].

\bibitem{roy}
  R.~Maartens and E.~Majerotto,
  %``Observational constraints on self-accelerating cosmology,''
  Phys.\ Rev.\  D {\bf 74} (2006) 023004 
  [arXiv:astro-ph/0603353].
  %%CITATION = PHRVA,D74,023004;%%

\bibitem{eisenstein}
  D.~J.~Eisenstein {\it et al.}  [SDSS Collaboration],
  %``Detection of the Baryon Acoustic Peak in the Large-Scale Correlation
  %Function of SDSS Luminous Red Galaxies,''
  Astrophys.\ J.\  {\bf 633} (2005) 560 
  [arXiv:astro-ph/0501171].


\bibitem{Riess1998}
  A.~G.~Riess {\it et al.}  [Supernova Search Team Collaboration],
  %``Observational Evidence from Supernovae for an Accelerating Universe and a
  %Cosmological Constant,''
  Astron.\ J.\  {\bf 116} (1998) 1009
  [arXiv:astro-ph/9805201].


%\cite{Riess:2004nr}
\bibitem{Riess:2004nr}
  A.~G.~Riess {\it et al.}  [Supernova Search Team Collaboration],
  %``Type Ia Supernova Discoveries at z>1 From the Hubble Space Telescope:
  %Evidence for Past Deceleration and Constraints on Dark Energy Evolution,''
  Astrophys.\ J.\  {\bf 607} (2004) 665
  [arXiv:astro-ph/0402512].
  %%CITATION = ASTRO-PH 0402512;%%

%\cite{Riess:2006fw}
\bibitem{riess07}
  A.~G.~Riess {\it et al.},
  %``New Hubble Space Telescope Discoveries of Type Ia Supernovae at $z > 1$:
  %Narrowing Constraints on the Early Behavior of Dark Energy,''
 Astrophys.\ J.\   {\bf 659} (2007) 98 
[arXiv:astro-ph/0611572].
  %%CITATION = ASTRO-PH 0611572;%%

\bibitem{inflation}
For reviews see e.g. S.~Tsujikawa,
  %``Introductory review of cosmic inflation,''
  arXiv:hep-ph/0304257,
N.~Turok,
  %``A critical review of inflation,''
  Class.\ Quant.\ Grav.\  {\bf 19} (2002) 3449.

\bibitem{flat cmb}
C.~L.~Bennett {\it et al.}  [WMAP Collaboration],
  %``First Year Wilkinson Microwave Anisotropy Probe (WMAP) Observations:
  %Preliminary Maps and Basic Results,''
  Astrophys.\ J.\ Suppl.\  {\bf 148} (2003) 1
  [arXiv:astro-ph/0302207];
D.~N.~Spergel {\it et al.}  [WMAP Collaboration],
  %``First Year Wilkinson Microwave Anisotropy Probe (WMAP) Observations:
  %Determination of Cosmological Parameters,''
  Astrophys.\ J.\ Suppl.\  {\bf 148} (2003) 175
  [arXiv:astro-ph/0302209].





\bibitem{dgp}
  G.~R.~Dvali, G.~Gabadadze and M.~Porrati,
  %``4D gravity on a brane in 5D Minkowski space,''
  Phys.\ Lett.\ B {\bf 485} (2000) 208
  [arXiv:hep-th/0005016].
\bibitem{deffayet}
C.~Deffayet,
  %``Cosmology on a brane in Minkowski bulk,''
  Phys.\ Lett.\ B {\bf 502} (2001) 199
  [arXiv:hep-th/0010186].
%\bibitem{riess07}A.~G.~Riess {\it et al.},
  %``New Hubble Space Telescope Discoveries of Type Ia Supernovae at $z > 1$:
  %Narrowing Constraints on the Early Behavior of Dark Energy,''
%  arXiv:astro-ph/0611572.
\bibitem{ldgp}V.~Sahni and Y.~Shtanov,
  %``Braneworld models of dark energy,''
  JCAP {\bf 0311} (2003) 014
  [arXiv:astro-ph/0202346];
A.~Lue and G.~D.~Starkman,
  %``How a brane cosmological constant can trick us into thinking that W <
  %-1,''
  Phys.\ Rev.\ D {\bf 70} (2004) 101501
  [arXiv:astro-ph/0408246].


\bibitem{cp}M.~Chevallier and D.~Polarski,
  %``Accelerating universes with scaling dark matter,''
  Int.\ J.\ Mod.\ Phys.\ D {\bf 10} (2001) 213
  [arXiv:gr-qc/0009008].

\bibitem{linder}E.~V.~Linder,
  %``Exploring the expansion history of the universe,''
  Phys.\ Rev.\ Lett.\  {\bf 90} (2003) 091301
  [arXiv:astro-ph/0208512].


\bibitem{qdgp paper}
  L.~P.~Chimento, R.~Lazkoz, R.~Maartens and I.~Quiros,
  %``Crossing the phantom divide without phantom matter,''
  JCAP {\bf 0609} (2006) 004
  [arXiv:astro-ph/0605450].

\bibitem{similar_papers}
  P.~Serra, A.~Heavens and A.~Melchiorri,
  %``Bayesian Evidence for a Cosmological Constant using new High-Redshift
  %Supernovae Data,''
  arXiv:astro-ph/0701338,
  %%CITATION = ASTRO-PH/0701338;%%
  V.~Barger, Y.~Gao and D.~Marfatia,
  %``Accelerating cosmologies tested by distance measures,''
  Phys.\ Lett.\  B {\bf 648} (2007) 127
  [arXiv:astro-ph/0611775],
  %%CITATION = PHLTA,B648,127;%%
  E.~L.~Wright,
  %``Constraints on Dark Energy from Supernovae, Gamma Ray Bursts, Acoustic
  %Oscillations, Nucleosynthesis and Large Scale Structure and the Hubble
  %constant,''
  arXiv:astro-ph/0701584.
  %%CITATION = ASTRO-PH/0701584;%%


\bibitem{dgp_goobar}
  S.~Rydbeck, M.~Fairbairn and A.~Goobar,
  %``Testing the DGP model with ESSENCE,''
  arXiv:astro-ph/0701495.
  %%CITATION = ASTRO-PH/0701495;%%


\bibitem{other_constraints}
  G.~B.~Zhao, J.~Q.~Xia, H.~Li, C.~Tao, J.~M.~Virey, Z.~H.~Zhu and X.~Zhang,
  %``Probing for dynamics of dark energy and curvature of universe with latest
  %cosmological observations,''
  Phys.\ Lett.\  B {\bf 648} (2007) 8
  [arXiv:astro-ph/0612728];
  %%CITATION = PHLTA,B648,8;%%
  N.~Pires, Z.~H.~Zhu and J.~S.~Alca\~niz,
  %``Lookback time as a test for brane cosmology,''
  Phys.\ Rev.\  D {\bf 73} (2006) 123530
  [arXiv:astro-ph/0606689];
  %%CITATION = PHRVA,D73,123530;%%
C.~Zunckel and R.~Trotta,
  %``Reconstructing the history of dark energy using maximum entropy,''
  arXiv:astro-ph/0702695;
  %%CITATION = ASTRO-PH/0702695;%%
  Y.~Gong, A.~Wang, Q.~Wu and Y.~Z.~Zhang,
  %``Direct evidence of acceleration from distance modulus redshift graph,''
  arXiv:astro-ph/0703583.
  %%CITATION = ASTRO-PH/0703583;%%


\bibitem{sahni}V.~Sahni and A.~Starobinsky,
  %``Reconstructing dark energy,''
  V.~Sahni and A.~Starobinsky,
  %``Reconstructing dark energy,''
  Int.\ J.\ Mod.\ Phys.\  D {\bf 15} (2006) 2105
  [arXiv:astro-ph/0610026].


\bibitem{dl}
A.~A.~Starobinsky,
  %``How to determine an effective potential for a variable cosmological
  %term,''
  JETP Lett.\  {\bf 68} (1998) 757
  [Pisma Zh.\ Eksp.\ Teor.\ Fiz.\  {\bf 68} (1998) 721
  [arXiv:astro-ph/9810431]; D.~Huterer and M.~S.~Turner,
  %``Revealing Quintessence,''
  Phys.\ Rev.\ D {\bf 60} (1999) 081301
  [arXiv:astro-ph/9808133];
  T.~D.~Saini, S.~Raychaudhury, V.~Sahni and A.~A.~Starobinsky,
  %``Reconstructing the Cosmic Equation of State from Supernova distances,''
  Phys.\ Rev.\ Lett.\  {\bf 85} (2000) 1162
  [arXiv:astro-ph/9910231]; T.~Chiba and T.~Nakamura,
  %``Feasibility of reconstructing the quintessential potential using SNIa
  %data,''
  Phys.\ Rev.\ D {\bf 62} (2000) 121301
  [arXiv:astro-ph/0008175];
  D.~A.~Dicus and W.~W.~Repko,
  %``Modeling dark energy with a top-down approach,''
  arXiv:astro-ph/0605537.
  %%CITATION = ASTRO-PH 0605537;%%

%\cite{Lazkoz:2005sp}
\bibitem{perivol}
  R.~Lazkoz, S.~Nesseris and L.~Perivolaropoulos,
%   ``Exploring Cosmological Expansion Parametrizations with the Gold SnIa
  %Dataset,''
  JCAP {\bf 0511} (2005) 010 
  [arXiv:astro-ph/0503230].
  %%CITATION = ASTRO-PH 0503230;%%

\bibitem{wans}
%\cite{Chevallier:2000qy}
%\bibitem{Chevallier:2000qy}
%  M.~Chevallier and D.~Polarski,
  %``Accelerating universes with scaling dark matter,''
%  Int.\ J.\ Mod.\ Phys.\ D {\bf 10} (2001) 213 
%  [arXiv:gr-qc/0009008];
  %%CITATION = GR-QC 0009008;%%
%%%
%\cite{Weller:2001gf}
%\bibitem{Weller:2001gf}
  J.~Weller and A.~Albrecht,
  %``Future Supernovae observations as a probe of dark energy,''
  Phys.\ Rev.\ D {\bf 65} (2002) 103512 
  [arXiv:astro-ph/0106079];
  %%CITATION = ASTRO-PH 0106079;%%
%%%%
%\cite{Gerke:2002sx}
%\bibitem{Gerke:2002sx}
  B.~F.~Gerke and G.~Efstathiou,
   %``Probing quintessence: Reconstruction and parameter estimation from
  %supernovae,''
  Mon.\ Not.\ Roy.\ Astron.\ Soc.\  {\bf 335} (2002) 33 
  [arXiv:astro-ph/0201336];
  %%CITATION = ASTRO-PH 0201336;%%
%%%%
%\cite{Maor:2001ku}
%\bibitem{Maor:2001ku}
  I.~Maor, R.~Brustein, J.~McMahon and P.~J.~Steinhardt,
  %``Measuring the Equation-of-state of the Universe: Pitfalls and Prospects,''
  Phys.\ Rev.\ D {\bf 65} (2002) 123003 
  [arXiv:astro-ph/0112526];
  %%CITATION = ASTRO-PH 0112526;%%
%%%%
%\cite{Corasaniti:2002vg}
%\bibitem{Corasaniti:2002vg}
  P.~S.~Corasaniti and E.~J.~Copeland,
  %``A model independent approach to the dark energy equation of state,''
  Phys.\ Rev.\ D {\bf 67}  (2003) 063521 
  [arXiv:astro-ph/0205544];
  %%CITATION = ASTRO-PH 0205544;%%
%%%%%
%\cite{Linder:2002et}
%\bibitem{Linder:2002et}
  E.~V.~Linder,
  %``Exploring the expansion history of the universe,''
  Phys.\ Rev.\ Lett.\  {\bf 90}  (2003) 091301 
  [arXiv:astro-ph/0208512];
  %%CITATION = ASTRO-PH 0208512;%%
%%%
%\cite{Wang:2003gz}
%\bibitem{Wang:2003gz}
  Y.~Wang and P.~Mukherjee,
%   ``Model-Independent Constraints on Dark Energy Density from Flux-averaging
  %Analysis of Type Ia Supernova Data,''
  Astrophys.\ J.\  {\bf 606} (2004) 654 
  [arXiv:astro-ph/0312192];
  %%CITATION = ASTRO-PH 0312192;%%
%%%%%
%\cite{Saini:2003wq}
%\bibitem{Saini:2003wq}
  T.~D.~Saini, J.~Weller and S.~L.~Bridle,
  %``Revealing the Nature of Dark Energy Using Bayesian Evidence,''
  Mon.\ Not.\ Roy.\ Astron.\ Soc.\  {\bf 348} (2004) 603 
  [arXiv:astro-ph/0305526];
  %%CITATION = ASTRO-PH 0305526;%%
%%%%
%\cite{Nesseris:2004wj}
%\bibitem{Nesseris:2004wj}
  S.~Nesseris and L.~Perivolaropoulos,
  %``A comparison of cosmological models using recent supernova data,''
  Phys.\ Rev.\ D {\bf 70} (2004) 043531 
  [arXiv:astro-ph/0401556];
  %%CITATION = ASTRO-PH 0401556;%%
%%%%
%\cite{Nesseris:2005ur}
%\bibitem{Nesseris:2005ur}
  S.~Nesseris and L.~Perivolaropoulos,
 %  ``Comparison of the Legacy and Gold SnIa Dataset Constraints on Dark Energy
  %Models,''
  Phys.\ Rev.\ D {\bf 72} (2005) 123519 
  [arXiv:astro-ph/0511040];
  %%CITATION = ASTRO-PH 0511040;%%
%%%%%
%\cite{Gong:2004ns}
%\bibitem{Gong:2004ns}
  Y.~G.~Gong,
  %``Supernova constraints on dark energy model,''
  Int.\ J.\ Mod.\ Phys.\ D {\bf 14} (2005) 599 
  [arXiv:astro-ph/0401207];
  %%CITATION = ASTRO-PH 0401207;%%
%%%%%%
%\cite{Gong:2004sd}
%\bibitem{Gong:2004sd}
  Y.~G.~Gong,
  %``Model independent analysis of dark energy I: Supernova fitting result,''
  Class.\ Quant.\ Grav.\  {\bf 22} (2005) 2121 
  [arXiv:astro-ph/0405446];
  %%CITATION = ASTRO-PH 0405446;%%
%%%
%%%
%\cite{Jassal:2004ej}
%\bibitem{Jassal:2004ej}
  H.~K.~Jassal, J.~S.~Bagla and T.~Padmanabhan,
  %``WMAP constraints on low redshift evolution of dark energy,''
  Mon.\ Not.\ Roy.\ Astron.\ Soc.\  {\bf 356} (2005) L11
  [arXiv:astro-ph/0404378];
  %%CITATION = ASTRO-PH 0404378;%%
%%%%
%\cite{Feng:2004ff}
%\bibitem{Feng:2004ff}
  B.~Feng, M.~Li, Y.~S.~Piao and X.~Zhang,
  %``Oscillating quintom and the recurrent universe,''
  Phys.\ Lett.\ B {\bf 634} (2006) 101 
  [arXiv:astro-ph/0407432];
  %%CITATION = ASTRO-PH 0407432;%%
%\cite{Uzan:2006mf}
%\bibitem{Uzan:2006mf}
  J.~P.~Uzan,
  %``The acceleration of the universe and the physics behind it,''
  Gen.\ Rel.\ Grav.\  {\bf 39} (2007) 307 
  [arXiv:astro-ph/0605313].
  %%CITATION = ASTRO-PH 0605313;%%


\bibitem{Nesseris:2005ur}
  S.~Nesseris and L.~Perivolaropoulos,
%  ``Comparison of the Legacy and Gold SnIa Dataset Constraints on Dark Energy
  %Models,''
  Phys.\ Rev.\ D {\bf 72} (2005) 123519
  [arXiv:astro-ph/0511040].



\bibitem{geom}
V.~Sahni, T.~D.~Saini, A.~A.~Starobinsky and U.~Alam,
  %``Statefinder -- a new geometrical diagnostic of dark energy,''
  JETP Lett.\  {\bf 77} (2003) 201 
  [Pisma Zh.\ Eksp.\ Teor.\ Fiz.\  {\bf 77}, 249 (2003)]
  [arXiv:astro-ph/0201498];
U.~Alam, V.~Sahni, T.~D.~Saini and A.~A.~Starobinsky,
  %``Exploring the Expanding Universe and Dark Energy using the Statefinder
  %Diagnostic,''
  Mon.\ Not.\ Roy.\ Astron.\ Soc.\  {\bf 344} (2003) 1057
  [arXiv:astro-ph/0303009].
U.~Alam, V.~Sahni, T.~D.~Saini and A.~A.~Starobinsky,
  %``Is there Supernova Evidence for Dark Energy Metamorphosis ?,''
  Mon.\ Not.\ Roy.\ Astron.\ Soc.\  {\bf 354} (2004) 275 
  [arXiv:astro-ph/0311364];
U.~Alam, V.~Sahni and A.~A.~Starobinsky,
  %``The case for dynamical dark energy revisited,''
  JCAP {\bf 0406} (2004) 008 
  [arXiv:astro-ph/0403687].


\bibitem{lazmaamaj}
R.~Lazkoz, R.~Maartens and E.~Majerotto,
  %``Observational constraints on phantom-like braneworld cosmologies,''
  Phys.\ Rev.\ D {\bf 74} (2006) 083510
  [arXiv:astro-ph/0605701].

\bibitem{jimenez_loeb}
  R.~Jimenez and A.~Loeb,
  %``Constraining Cosmological Parameters Based on Relative Galaxy Ages,''
  Astrophys.\ J.\  {\bf 573} (2002) 37
  [arXiv:astro-ph/0106145].
  %%CITATION = ASJOA,573,37;%%



\bibitem{stern}
R.~Jimenez, L.~Verde, T.~Treu and D.~Stern,
  %``Constraints on the equation of state of dark energy and the Hubble constant
  %from stellar ages and the CMB,''
  Astrophys.\ J.\  {\bf 593}  (2003) 622
  [arXiv:astro-ph/0302560].



\bibitem{gdds}R.~G.~Abraham {\it et al.},
  %``The Gemini Deep Deep Survey: I. Introduction to the Survey, Catalogs and
  %Composite Spectra,''
  Astron.\ J.\  {\bf 127} (2004) 2455
  [arXiv:astro-ph/0402436].
\bibitem{archival}  
 J.~Dunlop, J.~Peacock, H.~Spinrad, A.~Dey, R.~Jimenez, D.~Stern and R.~Windhorst,
  %``A 3.5 - Gyr - old galaxy at redshift 1.55,''
  Nature {\bf 381} (1996) 581;
  H.~Spinrad, A.~Dey, D.~Stern, J.~Dunlop, J.~Peacock, R.~Jimenez and R.~Windhorst,
  %``LBDS 53W091: An Old red Galaxy at z=1.552,''
  Astrophys.\ J. {\bf 484} (1997) 581
  [arXiv:astro-ph/9702233];
T.~Treu, M.~Stiavelli,  S.~Casertano, P.~Moller,and G.~Bertin,
  Mon.\ Not.\ Roy.\ Astron.\ Soc.\  {\bf 308} (1999) 1037
  [arXiv:astro-ph/9904327];
 T.~Treu, M.~Stiavelli, P.~Moller, S.~Casertano and G.~Bertin,
  %``The properties of field elliptical galaxies at intermediate redshift. II:
  %photometry and spectroscopy of an HST selected sample,''
  Mon.\ Not.\ Roy.\ Astron.\ Soc.\  {\bf 326} (2001) 221
   [arXiv:astro-ph/0104177];
     T.~Treu, M.~Stiavelli, S.~Casertano, P.~Moller,  and G.~Bertin,
  %``The properties of field elliptical galaxies at intermediate redshift. II:
  %photometry and spectroscopy of an HST selected sample,''
 Astrophys.\ J.\ Lett.\  {\bf 564} (2002) L13
  [arXiv:astro-ph/0111504];
   L.~A.~Nolan, J.~S.~Dunlop, R.~Jim\'enez and A.~F.~Heavens,
  %``F stars, metallicity, and the ages of red galaxies at z > 1,''
  Mon.\ Not.\ Roy.\ Astron.\ Soc.\  {\bf 341}, 464 (2003)
  [arXiv:astro-ph/0103450].




\bibitem{dunlop}
 R.~Jimenez, J.~MacDonald, J.~S.~Dunlop, P.~Padoan and J.~A.~Peacock,
  %``Synthetic stellar populations: single stellar populations, stellar interior
  %models and primordial proto-galaxies,''
  Mon.\ Not.\ Roy.\ Astron.\ Soc.\  {\bf 349} (2004) 240 
  [arXiv:astro-ph/0402271].
\bibitem{rob}
  S.~L.~Bridle, R.~Crittenden, A.~Melchiorri, M.~P.~Hobson, R.~Kneissl and A.~N.~Lasenby,
  %``Analytic marginalization over CMB calibration and beam uncertainty,''
  Mon.\ Not.\ Roy.\ Astron.\ Soc.\  {\bf 335} (2002) 1193 
  [arXiv:astro-ph/0112114].

\bibitem{bonamente}
  M.~Bonamente, M.~K.~Joy, S.~J.~La Roque, J.~E.~Carlstrom, E.~D.~Reese and K.~S.~Dawson,
  %``Measurement of the cosmic distance scale from Chandra X-ray imaging and
  %Sunyaev-Zel'dovich Effect mapping of high redshift clusters of galaxies,''
 Astrophys.\ J.\  {\bf 647} (2006) 25
[arXiv:astro-ph/0512349]


 
\bibitem{freedman}
  W.~L.~Freedman {\it et al.},
  %``Final Results from the Hubble Space Telescope Key Project to Measure the
  %Hubble Constant,''
  Astrophys.\ J.\  {\bf 553} (2001) 47
  [arXiv:astro-ph/0012376].


\bibitem{Wood-Vasey}
  W.~M.~Wood-Vasey {\it et al.},
  %``Observational Constraints on the Nature of the Dark Energy: First
  %Cosmological Results from the ESSENCE Supernova Survey,''
  arXiv:astro-ph/0701041.




%\cite{Astier:2005qq}
\bibitem{Astier}
  P.~Astier {\it et al.},
  %``The Supernova Legacy Survey: Measurement of Omega_M, Omega_Lambda and w
  %from the First Year Data Set,''
  Astron.\ Astrophys.\  {\bf 447} (2006) 31
  [arXiv:astro-ph/0510447].
  %%CITATION = ASTRO-PH 0510447;%%

\bibitem{Hamuy}
  M.~Hamuy, M.~M.~Phillips, N.~B.~Suntzeff, R.~A.~Schommer and J.~Maza,
  %``BVRI Light Curves for 29 Type Ia Supernovae,''
Astron.\ Jour.\ {\bf 112} (1996) 2408 
  [arXiv:astro-ph/9609064]




\bibitem{Jha}
S.~Jha, A.~G.~Riess and R.~P.~Kirshner,
  %``Improved Distances to Type Ia Supernovae with Multicolor Light Curve
  %Shapes: MLCS2k2,''
  Astrophys.\ J.\  {\bf 659} (2007) 122
  [arXiv:astro-ph/0612666].

\bibitem{goobar} A. Goobar, 2006 private communication.

\bibitem{min}
  E.~Di Pietro and J.~F.~Claeskens,
  %``Quintessence models faced with future supernovae data,''
  Mon.\ Not.\ Roy.\ Astron.\ Soc.\  {\bf 341} (2003) 1299
  [arXiv:astro-ph/0207332]; S.~Nesseris and L.~Perivolaropoulos,
  %``A comparison of cosmological models using recent supernova data,''
  Phys.\ Rev.\ D {\bf 70} (2004) 043531
  [arXiv:astro-ph/0401556];
  O.~Elgaroy and T.~Multam\"aki,
  %``Bayesian analysis of Friedmannless cosmologies,''
  JCAP {\bf 0609} (2006) 002
  [arXiv:astro-ph/0603053].



\bibitem{bet}
 J.~R.~Bond, G.~Efstathiou and M.~Tegmark,
  %``Forecasting Cosmic Parameter Errors from Microwave Background Anisotropy
  %Experiments,''
  Mon.\ Not.\ Roy.\ Astron.\ Soc.\  {\bf 291} (1997) L33
  [arXiv:astro-ph/9702100].

\bibitem{Percival}
  W.~J.~Percival {\it et al.}  [The 2dFGRS Team Collaboration],
  %``Parameter constraints for flat cosmologies from CMB and 2dFGRS power
  %spectra,''
  Mon.\ Not.\ Roy.\ Astron.\ Soc.\  {\bf 337} (2002) 1068
  [arXiv:astro-ph/0206256].

\bibitem{pernes}
  S.~Nesseris and L.~Perivolaropoulos,
  %``Crossing the phantom divide: Theoretical implications and observational
  %status,''
  JCAP {\bf 0701} (2007) 018
  [arXiv:astro-ph/0610092].

\bibitem{multamaki}
  O.~Elgaroy and T.~Multam\"aki,
  %``On using the CMB shift parameter in tests of models of dark energy,''
  arXiv:astro-ph/0702343.
\bibitem{wright}
  E.~L.~Wright,
  %``Constraints on Dark Energy from Supernovae, Gamma Ray Bursts, Acoustic
  %Oscillations, Nucleosynthesis and Large Scale Structure and the Hubble
  %constant,''
  arXiv:astro-ph/0701584.

\bibitem{wang07}
  Y.~Wang and P.~Mukherjee,
  %``Observational Constraints on Dark Energy and Cosmic Curvature,''
  arXiv:astro-ph/0703780.

 \bibitem{jussi}J. V\"aliviita, PhD thesis, Helsinki Institute of Physics (2005).
  \bibitem{trotta1}R. Trotta, PhD thesis, University of Geneva (2004).
\bibitem{trotta2}R. Trotta, Mon. Not. R. Astron. Soc., {\bf 378} (2007) 72. 
  \bibitem{trotta3}R. Trotta, 2007 private communication.
   \bibitem{cousins}R.D. Cousins, Am.\ J.\ Phys.\ {\bf 63} (1995) 5.
  
\bibitem{zech}G.~Zech,
  %``Confronting classical and Bayesian confidence limits to examples,''
  arXiv:hep-ex/0004011.

\bibitem{jussitalk}J. V\"aliviita, 2006 private communication.

 \bibitem{holmes}D.G. T. Denison, C.C. Holmes, B.K. Mallick, A.F.M. Smith, {\it Bayesian Methods for Nonlinear Classification and Regression (Wiley Series in Probability and Statistics)},
   (John Wiley \& Sons, 2004). 

\bibitem{liddle}
  A.~R.~Liddle, P.~Mukherjee and D.~Parkinson, Astron.\, Geophys. {\bf 47} (2006) 4.30.
  [arXiv:astro-ph/0608184].

\bibitem{Jeff}H. Jeffreys, {\it The Theory of Probability (Oxford Classic Texts in the Physical Sciences)} (Oxford University Press, 3rd. ed., 1998)

\bibitem{Wass} L. Wasserman,  Carnegie Mellon University Statistics Department Technical Report No. 666,  1997.



%%%%%%%%%%%%%%%%%%%%%%%%%%%%%%%%%%%%%%%%%%%%%%%%%%%%%%%%%%%%%%%%%%%%%%%%%%%%%%%%%%%%%%%%%%%%%%%%%%%%%%%%%%%%%%%%%%%%%%%%%%%%%%%%%%%%%%%%%%%%%%%%%%%%%%%%%%%%%%%%%%%%%%%%%%%%%%%



  \end{thebibliography}
\end{document}